\documentclass[12pt,a4paper]{article}
\usepackage[T2A]{fontenc}
\usepackage[cp1251]{inputenc}
\usepackage{cite}
\usepackage{mathtext}
\usepackage{amssymb,amsthm,amsmath}
\usepackage{array}
\usepackage[dvips]{epsfig}
\usepackage[russian, english]{babel}
\usepackage[title,titletoc]{appendix}

\begin{document}
\author{\bf Yu.A. Markov$\!\,$\thanks{e-mail:markov@icc.ru}
, M.A. Markova$\!\,$\thanks{e-mail:markova@icc.ru}}
\title{\Large \bf Mapping between the classical and pseudoclassical\\
models of a relativistic spinning particle in external\\
bosonic and fermionic fields. II}
%
%
\date{\normalsize \it Institute for System Dynamics and Control Theory Siberian Branch\\
of Academy of Sciences of Russia, P.O. Box 1233, 664033 Irkutsk, Russia}
\thispagestyle{empty}
\maketitle{}
\def\theequation{\arabic{section}.\arabic{equation}}

{
\noindent
{\bf Abstract.} The exact solution of a system of bilinear identities derived in the first part of our  work \cite{markov_part_I} for the case of real Grassmann-odd tensor aggregate of the type
$(S, V_{\mu}, \!\,^{\ast}T_{\mu \nu}, A_{\mu}, P)$ is obtained. The consistency of the solution with a corresponding system of bilinear identities including both the tensor variables and their derivatives $(\dot{S}, \dot{V}_{\mu},\!\,^{\ast}\dot{T}_{\mu \nu}, \dot{A}_{\mu}, \dot{P})$ is considered. The alternative approach in solving of the algebraic system based on introducing complex tensor quantities is discussed. This solution is used in constructing the mapping of the interaction terms of spinning particle with a background (Majorana) fermion field $\Psi^{i}_{{\rm M}\hspace{0.02cm}\alpha}(x)$.
A way of the extension of the obtained results for the case of  the Dirac spinors
$(\psi_{{\rm D}\hspace{0.03cm}\alpha}, \theta_{{\rm D}\hspace{0.03cm}\alpha})$ and a background Dirac field $\Psi^{i}_{{\rm D}\hspace{0.03cm}\alpha}(x)$, is suggested. It is shown that for the construction of one-to-one correspondence between the most general spinors and the tensor variables, we need a four-fold increase of the number of the tensor ones. A connection with the higher-order derivative Lagrangians for a point particle and in particular, with the Lagrangian suggested by A.M. Polyakov, is proposed.
}
{}

\newpage


\section{Introduction}
\setcounter{equation}{0}
\label{section_1}

In the second part of our work we proceed with our analysis of the problem of constructing an mapping between two Lagrangian descriptions of the spin degrees of freedom of a color spinning massive particle interacting with background non-Abelian gauge and fermion fields, started in \cite{markov_part_I} (to be referred to as “Paper I” throughout this text). Here, we will confine our attention to the interaction of the particle with background fermion field.
\\
\indent
In our considerations in Paper I so far we have dealt only with mapping bilinear combinations in the form $\bar{\psi} \hat{O} \psi$, where $\hat{O}$ is a certain (differential) operator or matrix, to the quadratic combinations $\xi_{\mu\,}\xi_{\nu},\, \xi_{\mu\,} \dot{\xi^{\mu}}$ etc. In our papers \cite{markov_J_Phys_G_2010, markov_NPA_2007, markov_IJMPA_2010} the simplest model classical Lagrangian of the interaction of a color spinning particle with an external non-Abelian fermion field $\Psi_{\alpha}^{i}(x)$ has been suggested. This Lagrangian has the following form:
\begin{equation}
L_{\Psi}=-\hspace{0.01cm}e g \hbar\hspace{0.04cm} (\bar{\theta}\theta) \sqrt{\frac{m}{2}}
\hspace{0.03cm}\Bigl\{\theta^{\dagger{i}}\bigl(\bar{\psi}_{\alpha}\Psi^{i}_{\alpha}(x)\bigr)
+ \bigl(\bar{\Psi}^{i}_{\alpha}(x)\hspace{0.03cm}\psi_{\alpha}\bigr)\hspace{0.03cm}\theta^{i}\Bigr\}
\;+
\label{eq:5q}
\end{equation}
\[
+\;\hspace{0.01cm}e g \hbar\hspace{0.04cm} (\bar{\theta}\theta) \sqrt{\frac{m}{2}} \biggl(\frac{C_F}{2\hspace{0.02cm}T_F}\biggr) Q^{a}\Bigl\{\theta^{\dagger{j}}(t^a)^{ji}\bigl(\bar{\psi}_{\alpha}\Psi^{i}_{\alpha}(x)\bigr)
+ \bigl(\bar{\Psi}^{i}_{\alpha}(x)\hspace{0.03cm}\psi_{\alpha}\bigr)(t^a)^{ij}\theta^{j}\Bigr\}.
\]
Here, in contrast to \cite{markov_J_Phys_G_2010}, we have separated out in an explicit form the dependence of $\hbar$, introduced the dimensional\footnote{\,In \cite{markov_J_Phys_G_2010} the commutative spinor $\psi_{\alpha}$ was considered as a dimensional quantity with the dimension  [mass]$^{1/2}$. In the previous and present works the $\psi_{\alpha}$ spinor is dimensionless.} factor $m^{1/2}$ and dimensionless nilpotent one $(\bar{\theta} \theta) \equiv \bar{\theta}_{\alpha} \theta_{\alpha}$. It is easy to see that the Lagrangian has the proper dimension for the canonical dimension of an external fermion field (in units $c=1$)
\[
\bigl[\Psi^{i}_{\alpha}(x)\bigr] \sim \frac{1}{\;[\,{\rm time}\,]^{3/2}}.
\]
The commuting spinor $\psi_{\alpha}(\tau)$ enters linearly into the expression (\ref{eq:5q}). This spinor is connected with a set of the anticommuting tensor quantities $(S, V_{\mu}, \!\,^{\ast}T_{\mu\nu},\ldots)$ by means of the general relation (I.2.1) (references to formulas in \cite{markov_part_I} are prefixed by the roman number I), if we preliminarily contract the latter with the auxiliary spinor $\theta_{\beta}$
\begin{equation}
\hbar^{1/2}(\bar{\theta}\theta)\psi_{\alpha} = \frac{1}{4}\,
\Bigl\{-i\hspace{0.01cm}S\hspace{0.02cm}\theta_{\alpha} + V_{\mu}(\gamma^{\mu}\theta)_{\alpha} - \frac{i}{2}\,^{\ast}T_{\mu\nu}(\sigma^{\mu\nu}\gamma_{5}\hspace{0.02cm}\theta)_{\alpha} +
i\hspace{0.01cm}A_{\mu}(\gamma^{\mu}\gamma_{5}\hspace{0.02cm}\theta)_{\alpha} + P(\gamma_{5}\hspace{0.03cm}\theta)_{\alpha}\Bigr\}.
\label{eq:5w}
\end{equation}
Such an approach of recovering a spinor from the Clifford algebra aggregate by means of an arbitrary auxiliary spinor (so-called the {\it inverse theorem}) was considered  in the commutative case in a number of papers: Zhelnorovich \cite{zhelnorovich_1972}, Crawford \cite{crawford_1985},
Keller and Rodriguez-Romo \cite{keller_1990}, Rodriguez-Romo \cite{rodriguez_1993}, Lounesto \cite{lounesto_1993}  (see also Klauder \cite{klauder_1964}). If we substitute the representation (\ref{eq:5w}) into the Lagrangian (\ref{eq:5q}), then in contrast to the mapping of the above-mentioned bilinear combinations, the auxiliary spinor $\theta_{\alpha}$ will already enter explicitly into the Lagrangian $L_{\Psi}$  as an independent entity. Here, a difficult and subtle question of the dependence of the mapped Lagrangian (\ref{eq:5q}) on a concrete choice of $\theta_{\alpha}$ arises, whether it is possible to give a physical meaning of the auxiliary spinor. A similar problem was discussed in paper \cite{zhelnorovich_1972}.\\
\indent
Furthermore, in the general expansion (\ref{eq:5w}) not all of the functions
$(S, V_{\mu},\!\,^{\ast}T_{\mu \nu},\ldots)$ are independent by virtue of the relations (I.C.1)\,--\,(I.C.15). Although we have already used some of these relations in the analysis of mapping the bilinear combinations, however, we have not explicitly resolved them. Solutions of these equations would define an explicit connection between the quantities in tensor set $(S, V_{\mu},\!\,^{\ast}T_{\mu \nu},\ldots)$ and ipso facto would reduce the number of quantities in (\ref{eq:5w}). Three subsequent sections are concerned with the problem.\\
\indent
Paper II is organized as follows. Sections \ref{section_5}, \ref{section_6} and \ref{section_7} are concerned with the construction of the explicit solutions for a system of bilinear algebraic equations defined in Section  2 and in Appendix C of Paper I. In Section  \ref{section_5} a well-known example of deriving the explicit solution for the {\it commutative} tensor variables $(S, V_{\mu}, \!\,^{\ast}T_{\mu \nu}, A_{\mu}, P)$ is reproduced. In Section \ref{section_6}  this analysis is extended to a qualitative new case of {\it anticommuting} tensor set. The exact solution of the corresponding system of equations is presented and also an analysis of solving a system of equations including the derivatives of the tensor variables, is given. Finally, in Section \ref{section_7} the alternative approach in solving the problem based on introducing {\it complex} tensor quantities, is considered. This enables us to provide also more insight into the structure of an algebraic system of identities and to see that the system under consideration in principle admits the existence of the second independent solution.\\
\indent
Section \ref{section_8} supplies a detailed discussion of the possibility of the existence of the second solution. Here we give conditions under which the second solution is permissible. It was shown that there exists the only algebraic identity which results in unremovable contradiction at least in a class of Majorana spinors.\\
\indent
In Section \ref{section_10}, one possible way of extension of the results obtained in the previous sections to a much more complicated case of the Dirac spinors $\psi_{{\rm D}\hspace{0.01cm}\alpha}$ and $\theta_{{\rm D}\hspace{0.01cm}\alpha}$, is suggested. On the basis of a general analysis it is concluded that for the existence of (one-to-one) correspondence between the Dirac spinors
$(\psi_{\rm D}, \theta_{\rm D})$ and tensor variables
$(S, V_{\mu}, \!\,^{\ast}T_{\mu \nu}, A_{\mu}, P)$  it is necessary a fourfold increase of the number of the real tensor ones $(S^{ij}, V_{\mu}^{ij},\!\,^{\ast}T_{\mu \nu}^{ij}, A_{\mu}^{ij}, P^{ij})$, where $i,j=1, 2$. \\
\indent
Section \ref{section_11} is concerned with a discussion of the possibility of a connection of the approach stated in the previous sections, with the models of a spinning particle based on the higher-order derivative Lagrangians.\\
\indent
In the concluding Section 8 we briefly discuss the construction of the mapping between the systems obtained after quantization of these classical models.\\
\indent
In Appendix \ref{appendix_E} the parameter representation of orthogonal tetrad $h_{\mu}^{(s)}$ used in deriving solutions of the algebraic equations, is given. In Appendix \ref{appendix_F} a way of the construction of a system of bilinear identities for the {\it complex} tensor variables $(S, V_{\mu}, \!\,^{\ast}T_{\mu \nu}, \ldots)$, is suggested. A system of identities in Appendix C of Paper\,I is a just special case.


\section{Solving the algebraic equations. The commutative case}
\setcounter{equation}{0}
\label{section_5}

\indent
It is rather useful at the beginning to deal with the case of {\it commuting} tensor variables, as was first considered in the papers by Takahashi {\it et al.} \cite{takahashi_1982, takahashi_1983}. In these papers three independent equations of a complete system were analyzed. In our case these equations\footnote{\,Our variables in the commutative case are connected with analogous those in paper \cite{takahashi_1983} by the relations
\[
S \rightarrow iJ, \,\; P \rightarrow -iJ_5, \,\; V_{\mu} \rightarrow J_{\mu}, \,\; A_{\mu} \rightarrow -iJ_{5\mu},
\,^{\ast}T_{\mu\nu} \rightarrow \,^{\ast}\!J_{\mu\nu}, \,\; T_{\mu\nu} \rightarrow J_{\mu\nu}.
\]
In addition, our definitions of the matrix $\sigma^{\mu\nu}$ and antisymmetric tensor $\epsilon^{\mu\nu\lambda\sigma}$ differ from those of \cite{takahashi_1983} by signs.}
follow from (I.C.14),  (I.C.4),  (I.C.6) and  (I.C.13)
\begin{align}
T^{\mu\nu\hspace{0.02cm}}V_{\nu} &= P\!\hspace{0.02cm}A^{\mu}, \label{eq:5e}\\
\,^{\ast}T^{\mu\nu\hspace{0.02cm}}V_{\nu} &= S\!\hspace{0.03cm}A^{\mu}, \label{eq:5r}\\
V_{\mu}V^{\mu} &= A_{\mu}A^{\mu} = -\hspace{0.01cm}(S^2 + P^2).\label{eq:5t}
\end{align}
Assuming $S\neq 0$, we obtain the pseudovector $A^{\mu}$ from equation (\ref{eq:5r}) and substitute it into (\ref{eq:5e})
\[
\bigl(S\hspace{0.03cm}T^{\mu\nu}\!- P\,^{\ast}T^{\mu\nu}\bigr) V_{\nu} = 0.
\]
Next we consider a particular representation of the tensor $T_{\mu\nu}$ and of its dual one in terms of two independent functions $S$ and $P$
\begin{equation}
\begin{split}
T^{\mu\nu} &=\, \omega^{\hspace{0.02cm}\mu\nu}\!\hspace{0.03cm}S + \,^{\ast}\omega^{\hspace{0.02cm}\mu\nu}\!\hspace{0.03cm}P,\\
\,^{\ast}T^{\mu\nu} &= \!\,^{\ast}\omega^{\hspace{0.02cm}\mu\nu}\!\hspace{0.03cm}S - \omega^{\hspace{0.02cm}\mu\nu}\!\hspace{0.03cm}P,
\end{split}
\label{eq:5y}
\end{equation}
where $\omega^{\mu\nu}$ is a certain antisymmetric tensor, an explicit form of which will be written just below. Substituting these expressions into the equation just above, we result in homogeneous algebraic equation for the vector variable $V_{\mu}$
\[
\omega^{\mu\nu}V_{\nu} = 0.
\]
A condition for the existence of a nontrivial solution ($\det\omega^{\mu\nu}=0$) places a restriction on the tensor $\omega_{\mu\nu}$
\begin{equation}
\omega_{\mu\nu}\!\,^{\ast}\omega^{\hspace{0.02cm}\mu\nu} = 0.
\label{eq:5u}
\end{equation}
One more restriction on the $\omega_{\mu\nu}$ tensor can be obtained if we consider the pseudoscalar equation  (I.C.5). In the commutative case it takes the form
\begin{equation}
S\!\hspace{0.04cm}P = \frac{1}{4}\,T_{\mu\nu}\!\,^{\ast}T^{\mu\nu}.
\label{eq:5i}
\end{equation}
Substituting the expressions (\ref{eq:5y}) into this equation, we result in the relation
\[
S\!\hspace{0.04cm}P = \frac{1}{4}\,\omega_{\mu\nu}\!\,^{\ast}\omega^{\hspace{0.02cm}\mu\nu}(S^2 - P^2) - \frac{1}{2}\,\omega_{\hspace{0.02cm}\mu\nu}\hspace{0.04cm}\omega^{\mu\nu}SP.
\]
To convert this relation into the identity, it is necessary to add to the condition (\ref{eq:5u}) another condition
\begin{equation}
\omega_{\mu\nu}\,\omega^{\hspace{0.02cm}\mu\nu} = -\hspace{0.01cm}2.
\label{eq:5o}
\end{equation}
In papers \cite{takahashi_1982, takahashi_1983} the parametrical solution of algebraic system (\ref{eq:5e})\,--\,(\ref{eq:5t}) was given in terms of the Euler angles $\alpha,\,\beta,\,\gamma$ and pseudoangles $\chi_i,\, i=1, 2, 3$ for a Lorentz boost. Following these papers, if we express the original commuting spinor $\psi_{\alpha}$ in terms of the same parameters $(\alpha, \beta, \gamma, \chi_i)$, then we ipso facto specify a {\it parametrical} connection between the spinor and the tensor set $(S,\,V_{\mu},\,^{\ast}T_{\mu\nu},\,A_{\mu},\,P)$ (see Crawford \cite{crawford_1985}). In the case of (\ref{eq:5w}) at the cost of introducing the auxiliary spinor $\theta_{\alpha}$ such a connection is more direct.\\
\indent
A quite compact and explicit representation of solutions for the system (\ref{eq:5e})\,--\,(\ref{eq:5t}) is obtained by using so-called tetrad $\bigl(h_{\mu}^{(s)}\bigr), \,s= 0, 1, 2, 3$, i.e. a set of four linear independent orthogonal unit 4-vectors $h_{\mu}^{(s)}$ (numbered by the index $s$) subject to the relations
\begin{equation}
\begin{split}
&h_{\mu}^{(s)}h^{(s^{\prime})\mu} = g^{ss^{\prime}} = {\rm diag}(1,-1,-1,-1),\\
&h_{\mu}^{(s)}h_{\nu}^{(s^{\prime})}g_{ss^{\prime}} = g_{\mu\nu} = {\rm diag}(1,-1,-1,-1).
\end{split}
\label{eq:5p}
\end{equation}
An explicit form of the tetrad vectors is given in Appendix \ref{appendix_E}. These vectors (more exactly, a complete matrix of an arbitrary rotation of Minkowski space, in which the components of these four vectors are columns) have been first introduced into consideration in the remarkable monograph by J.L. Synge \cite{synge_book_1956}. In what follows we need only two vectors $h_{\mu}^{(1)}$ and $h_{\mu}^{(2)}$.\\
\indent
Let us define an antisymmetric tensor $\omega_{\mu\nu}$ in the following form:
\begin{equation}
\begin{split}
\omega_{\mu\nu} &= - \,\epsilon_{\mu\nu\lambda\sigma} h^{(1)\lambda}h^{(2)\sigma}
\;\bigl(\hspace{0.02cm} \equiv h_{\mu}^{(0)}h_{\nu}^{(3)} - h_{\nu}^{(0)}h_{\mu}^{(3)}\bigr),\\
\,^{\ast}\omega_{\mu\nu} &= h_{\mu}^{(1)}h_{\nu}^{(2)} - h_{\nu}^{(1)}h_{\mu}^{(2)}.
\end{split}
\label{eq:5a}
\end{equation}
It is easy to convince ourselves that the conditions (\ref{eq:5u}) and (\ref{eq:5o}) hold by virtue of the properties (\ref{eq:5p}). Taking into account the representation (\ref{eq:5y}), we obtain that the following functions (we consider that $S^{\,2} + P^{\,2}\! \neq 0$):
\begin{equation}
A_{\mu} = (S^{\,2} + P^{\,2})^{\!1/2} h^{(1)}_{\mu},
\qquad
V_{\mu} = -(S^{\,2} + P^{\,2})^{\!1/2} h^{(2)}_{\mu}
\label{eq:5s}
\end{equation}
satisfy the system (\ref{eq:5e})\,--\,(\ref{eq:5t}). It is interesting to note that the tensor $T_{\mu \nu}$ after elimination of 4-vectors $h_{\mu}^{(1)}$ and $h_{\mu}^{(2)}$ with the aid of (\ref{eq:5s}), can be presented solely in terms of $V_{\mu},A_{\mu},P \mbox{and}\,S$ \cite{crawford_1985}
\[
T_{\mu\nu} = -\frac{1}{(S^{\,2} + P^{\,2})}\,\bigl\{(A_{\mu}V_{\nu} - A_{\nu}V_{\mu})P - \epsilon_{\mu\nu\lambda\sigma}A^{\lambda}V^{\sigma\!}S\bigr\}.
\]
\indent
In closing this section it should be pointed out that the parametrization of solutions of the algebraic equations presented here, is not only possible. In the paper by Takabayasi \cite{takabayasi_1958} a different approach to the construction of an explicit form of real and mutually orthogonal tetrad $h_{\mu}^{(s)}$ was given. For constructing the tetrad in \cite{takabayasi_1958} an arbitrary commuting Dirac spinor $\varphi$ was used. In terms of the spinor four mutually orthogonal real 4-vectors are defined as follows:
\[
\begin{split}
&s_{\mu} \equiv \bar{\varphi}\hspace{0.03cm}\gamma_{\mu}\varphi, \quad a_{\mu} \equiv \bar{\varphi}\hspace{0.03cm}\gamma_{\mu}\gamma_{5}\varphi,\\
&\alpha^{(1)}_{\mu} \equiv \frac{1}{2}\,(\bar{\varphi}^c\gamma_{\mu}\varphi - \bar{\varphi}\hspace{0.03cm}\gamma_{\mu}\varphi^c),\\
&\alpha^{(2)}_{\mu} \equiv \frac{1}{2}\,(\bar{\varphi}^c\gamma_{\mu}\varphi + \bar{\varphi}\hspace{0.03cm}\gamma_{\mu}\varphi^c),\\
\end{split}
\]
where $\varphi^c$ denotes the charge conjugated spinor: $\varphi^c = C\bar{\varphi}^T$. The required tetrad $h_{\mu}^{(s)}$ is connected with these vectors by the simple relations
\begin{equation}
\varrho\hspace{0.02cm} h^{(0)}_{\mu} \equiv s_{\mu},
\quad
\varrho\hspace{0.02cm} h^{(1,\,2)}_{\mu} \equiv \alpha^{(1,\,2)}_{\mu},
\quad
\varrho\hspace{0.02cm} h^{(3)}_{\mu} \equiv a_{\mu},
\label{eq:5d}
\end{equation}
where $\varrho \equiv \bigl((\bar{\varphi}\varphi)^2 + (\bar{\varphi}\gamma_{5}\varphi)^2 \bigr)^{1/2}$. Further, in the papers by Nash \cite{nash_1980} one more approach to the construction of the tetrad by a real eight-component $\overline{\rm O(3,3)}$ spinor is presented.\\
\indent
Finally, we will also discuss another alternative presentation of tetrad $h_{\mu}^{(s)}$ in Section \ref{section_11}.


\section{\!\!\!Solving the algebraic equations. \!\!The anticommutative case}
\setcounter{equation}{0}
\label{section_6}

Now we turn our attention to solving the algebraic equations for the case of anticommuting quantities $(S, V_{\mu},\!\,^{\ast}T_{\mu\nu}\ldots )$. By virtue of nilpotency of the quantities, equation (\ref{eq:5t})
vanishes. The following equations
\begin{equation}
\begin{split}
T^{\mu\nu}V_{\nu} &= -P\!\hspace{0.02cm}A^{\mu} + 2\hspace{0.02cm} S\hspace{0.02cm}V^{\mu} ,\\
\,^{\ast}T^{\mu\nu}V_{\nu} &= -S\!\hspace{0.02cm}A^{\mu} -  2\hspace{0.02cm}P\hspace{0.02cm}V^{\mu}
\end{split}
\label{eq:6q}
\end{equation}
are analogous to equations (\ref{eq:5e}) and (\ref{eq:5r}), correspondingly. The first of them follows from (I.C.14), and the second does from (I.C.4). The right-hand sides of these equations have a somewhat more complicated structure in comparison with the commutative case (\ref{eq:5e}), (\ref{eq:5r}) and one would expect that it will be reflected in the structure of their solutions.\\
\indent
Let us present the anticommuting tensor quantity $T^{\mu \nu}$ (and its dual one) in the form similar to (\ref{eq:5y})
\begin{equation}
\begin{split}
T^{\mu\nu} &=\, \omega^{\mu\nu}\!\hspace{0.02cm}S + \,^{\ast}\omega^{\mu\nu}\!\hspace{0.02cm}P,\\
\,^{\ast}T^{\mu\nu} &= \,^{\ast}\omega^{\mu\nu}\!\hspace{0.02cm}S - \omega^{\mu\nu}\!\hspace{0.02cm}P,
\end{split}
\label{eq:6w}
\end{equation}
where the commuting antisymmetric tensor $\omega_{\mu \nu}$ and its dual one are
\begin{equation}
\begin{split}
\omega_{\mu\nu} &= - \,\epsilon_{\mu\nu\lambda\sigma} h^{(1)\lambda}h^{(2)\sigma},\\
\,^{\ast}\omega_{\mu\nu} &= h_{\mu}^{(1)}h_{\nu}^{(2)} - h_{\nu}^{(1)}h_{\mu}^{(2)}.
\end{split}
\label{eq:6e}
\end{equation}
Recall that in (\ref{eq:6w}) the functions $P$ and $S$ are now nilpotent. It is not difficult to see by using the first property in the definition of tetrad (\ref{eq:5p}) that in the case under consideration the following functions are the solution of system (\ref{eq:6q})
\begin{equation}
\begin{split}
A_{\mu} &= P h^{(1)}_{\mu} + S h^{(2)}_{\mu},\\
V_{\mu} &= S h^{(1)}_{\mu} - P h^{(2)}_{\mu}.
\end{split}
\label{eq:6r}
\end{equation}
One can consider an alternative system of equations, ``dual'' to (\ref{eq:6q}), also having one external vector index,
\begin{equation}
\begin{split}
T^{\mu\nu\!}\!\hspace{0.02cm}A_{\nu} &= P\hspace{0.02cm}V^{\mu} + 2\hspace{0.02cm} S\!\hspace{0.03cm}A^{\mu} ,\\
\,^{\ast}T^{\mu\nu\!}\!\hspace{0.02cm}A_{\nu} &= S\hspace{0.02cm}V^{\mu} -  2\hspace{0.02cm}P\!\hspace{0.03cm}A^{\mu}
\end{split}
\label{eq:6t}
\end{equation}
(the first equation follows from (I.C.9), and the second one does from (I.C.2)) and verify that the solution (\ref{eq:6w})\,--\,(\ref{eq:6r}) also satisfies this system.\\
\indent
In Section 2 of Paper I we have obtained three independent equations with two external vector indices (I.2.12)\,--\,(I.2.14):
\begin{equation}
P\,^{\hspace{0.02cm}\ast}T^{\mu\nu} - S\hspace{0.04cm}T^{\mu\nu} = V^{\mu}V^{\nu} + A^{\mu\!}A^{\nu},
\label{eq:2a}
\end{equation}
\begin{equation}
P\,^{\hspace{0.01cm}\ast}T^{\mu\nu} + S\hspace{0.04cm}T^{\mu\nu} = -\,^{\ast}T^{\mu\lambda}\,^{\ast}T_{\lambda\;\;}^{\;\;\nu},
\hspace{0.45cm}
\label{eq:2s}
\end{equation}
\begin{equation}
-\hspace{0.02cm}\epsilon^{\hspace{0.02cm}\mu\nu\lambda\sigma}V_{\lambda}A_{\sigma} = V^{\mu}V^{\nu} - A^{\mu\!}A^{\nu}.
\hspace{0.5cm}
\label{eq:2d}
\end{equation}
By direct substitution of (\ref{eq:6w})\,--\,(\ref{eq:6r}) into (\ref{eq:2a}) we  get the identity
\[
-\,^{\ast}\omega^{\mu\nu\!}\hspace{0.02cm}S\!\hspace{0.04cm}P = -\,^{\ast}\omega^{\mu\nu\!}\hspace{0.02cm}S\!\hspace{0.04cm}P.
\]
Because of the structure of solution (\ref{eq:6w})\,--\,(\ref{eq:6r}) the following equalities hold:
\begin{equation}
V^{\mu}V^{\nu} = A^{\mu}A^{\nu}, \qquad  P\,^{\ast}T^{\mu\nu} = -S\hspace{0.03cm}T^{\mu\nu}.
\label{eq:6y}
\end{equation}
Further, due to the last equality in (\ref{eq:6y}) the left-hand side of equation (\ref{eq:2s}) is equal to zero and, in its turn, on the right-hand side we have
\[
\,^{\ast}T^{\mu\lambda}\,^{\ast}T_{\lambda\;\;}^{\;\;\nu} = \bigl(\,\omega^{\mu\lambda}\,^{\ast}\omega_{\lambda\;\;}^{\;\;\nu} - \,^{\ast}\omega^{\mu\lambda}\omega_{\lambda\;\;}^{\;\;\nu}\bigr)S\!\hspace{0.04cm}P.
\]
The right-hand side vanishes by virtue of the identity
\[
\,^{\ast}\omega^{\mu\lambda}\omega_{\lambda\;\;}^{\;\;\;\nu} = \omega^{\mu\lambda}\,^{\ast}\omega_{\lambda\;\;}^{\;\;\;\nu} = -\frac{1}{4}\bigl(\omega^{\lambda\sigma}\,^{\ast}\omega_{\lambda\sigma}\bigr)g^{\mu\nu},
\]
which holds for any antisymmetric tensor of second rank.\\
\indent
Finally, the right-hand side of equation (\ref{eq:2d}) vanishes by virtue of the first equality in (\ref{eq:6y}), and the left-hand side
\[
\epsilon^{\mu\nu\lambda\sigma}V_{\lambda}A_{\sigma} = (S^2 + P^2)\,\epsilon^{\mu\nu\lambda\sigma\hspace{0.01cm}}\,^{\ast}\omega_{\lambda\sigma}
+ S\!\hspace{0.04cm}P \epsilon^{\mu\nu\lambda\sigma}(h^{(1)}_{\lambda}h^{(1)}_{\sigma} +\, h^{(2)}_{\lambda}h^{(2)}_{\sigma})
\]
equals zero by nilpotency of $S$ and $P$, and by the antisymmetry of $\epsilon^{\mu \nu \lambda \sigma}$.\\
\indent
In addition to equations of the vector and tensor type, let us consider equation of the (pseudo)scalar type (I.C.5). In the anticommutative case it takes the form
\begin{equation}
S\!\hspace{0.04cm}P = -\frac{1}{2}\,V_{\mu}\hspace{0.02cm}A^{\mu}.
\label{eq:6u}
\end{equation}
Note that in the commutative case we have the condition $V_{\mu}\hspace{0.02cm}A^{\mu}=0$, as it follows from (I.C.8). It is easy to see that the solution (\ref{eq:6r}) satisfies equation (\ref{eq:6u}).\\
\indent
On examination of mapping the kinetic term in Section 3 of Paper I, we derived a certain algebraic relations between the basic quantities $(S, V_{\mu},\!\,^{\ast}T_{\mu\nu}\ldots)$ and their derivatives. In particular, we obtain two independent relations in the form
\begin{equation}
2\hspace{0.02cm}(S\!\hspace{0.04cm}\dot{S} + P\!\hspace{0.04cm}\dot{P})
= V_{\mu}\dot{V}^{\mu} + A_{\mu}\dot{A}^{\mu},
\label{eq:6i}
\end{equation}
\begin{equation}
3\hspace{0.02cm}(S\!\hspace{0.04cm}\dot{S} - P\!\hspace{0.04cm}\dot{P}) = - \frac{1}{2}\hspace{0.01cm}\,^{\ast}T_{\mu\nu}\!\,^{\ast}\dot{T}^{\mu\nu}.
\label{eq:6o}
\end{equation}
The question may now be raised whether the parametric solution (\ref{eq:6w})\,--\,(\ref{eq:6r}) identically satisfies equations (\ref{eq:6i}) and (\ref{eq:6o}) or additional restrictions can arise. Before we turn to this problem we shall supplement a system of the ``scalar'' equations (\ref{eq:6i}) and (\ref{eq:6o}) by two other independent equations of the ``pseudoscalar'' type. For this purpose we make use of equation (I.C.5) from the complete system in Appendix C of Paper I. By means of the rule described in Section 3 of Paper I, from (I.C.5) one can define two required equations of the pseudoscalar type
\begin{align}
&\dot{S}\!\hspace{0.05cm}P = \frac{1}{4}\,(S\!\hspace{0.05cm}\dot{P} + P\!\hspace{0.05cm}\dot{S}) -
\frac{1}{4}\,( V_{\mu}\hspace{0.02cm}\dot{A}^{\mu} - A_{\mu}\dot{V}^{\mu}) + \frac{1}{8}\,T_{\mu\nu\!}\,^{\ast}\dot{T}^{\mu\nu}, \notag\\
&\dot{P}\!\hspace{0.06cm}S = \frac{1}{4}\,(S\!\hspace{0.05cm}\dot{P} + P\!\hspace{0.05cm}\dot{S}) + \frac{1}{4}\,( V_{\mu}\hspace{0.02cm}\dot{A}^{\mu} - A_{\mu}\dot{V}^{\mu}) + \frac{1}{8}\,T_{\mu\nu}\!\,^{\ast}\dot{T}^{\mu\nu}. \notag
\end{align}
By adding and subtracting these two equations we obtain the desired system
\begin{align}
2\hspace{0.02cm}(P\!\hspace{0.05cm}\dot{S} - S\!\hspace{0.05cm}\dot{P}) &= V_{\mu\hspace{0.02cm}}\dot{A}^{\mu} - A_{\mu}\dot{V}^{\mu},
\label{eq:6p}\\
3\hspace{0.02cm}(P\!\hspace{0.05cm}\dot{S} + S\!\hspace{0.05cm}\dot{P}) &= -\frac{1}{2}\,T_{\mu\nu\!}\,^{\ast}\dot{T}^{\mu\nu}
\label{eq:6a}
\end{align}
in addition to equations (\ref{eq:6i}) and (\ref{eq:6o}).\\
\indent
Let us first look at equation (\ref{eq:6p}). Considering that the 4-vectors $h_{\mu}^{(1)}$ and $h_{\mu}^{(2)}$ in the general case are functions of the evolution parameter $\tau$ and using the presentation (\ref{eq:6r}), we obtain
\[
V_{\mu}\dot{A}^{\mu} = (P\!\hspace{0.05cm}\dot{S} - S\!\hspace{0.04cm}\dot{P})
+  S\!\hspace{0.04cm}P \hspace{0.02cm}\bigl(h^{(1)}_{\mu}\dot{h}^{(1)\mu} +\, h^{(2)}_{\mu}\dot{h}^{(2)\mu}\bigr).
\]
Here, the last term vanishes by virtue of the normalization $(h^{(1)})^2=(h^{(2)})^2=-1$. A similar expression with the opposite sign holds for $A_{\mu} \dot{V}^{\mu}$, i.e.
\[
A_{\mu}\dot{V}^{\mu} = -\hspace{0.02cm}(P\!\hspace{0.04cm}\dot{S} - S\!\hspace{0.04cm}\dot{P}).
\]
Thus, equation (\ref{eq:6p}) goes over into the identity.\\
\indent
Further, we consider equation (\ref{eq:6i}). For the presentation (\ref{eq:6r}) we have
\begin{equation}
V_{\mu}\dot{V}^{\mu} = A_{\mu}\dot{A}^{\mu} = -\hspace{0.02cm}(S\!\hspace{0.05cm}\dot{S} + P\!\hspace{0.04cm}\dot{P}) -
S\!\hspace{0.04cm}P (h^{(1)}_{\mu}\dot{h}^{(2)\mu} -\,\dot{h}^{(1)}_{\mu}h^{(2)\mu}).
\label{eq:6s}
\end{equation}
Here, however, the second term containing the derivatives of $h_{\mu}^{(1, 2)}$ in the general case does not vanish (the orthogonality condition $h_{\mu}^{(1)} h^{(2)\mu}=0$ involves the equality $h^{(1)}_{\mu}\dot{h}^{(2)\mu} + \dot{h}^{(1)}_{\mu}h^{(2)\mu}\!~=~0$). By this means equation (\ref{eq:6i}) after substituting (\ref{eq:6s}) and collecting similar terms takes the form of nontrivial relation between the derivatives of commuting and anticommuting quantities
\begin{equation}
2\hspace{0.02cm}(S\!\hspace{0.055cm}\dot{S} + P\!\hspace{0.045cm}\dot{P}) = -S\!\hspace{0.04cm}P \hspace{0.02cm}\bigl(h^{(1)}_{\mu}\dot{h}^{(2)\mu} -\,\dot{h}^{(1)}_{\mu}h^{(2)\mu}\bigr).
\label{eq:6d}
\end{equation}
\indent
Finally, let us consider two remaining equations (\ref{eq:6o}) and (\ref{eq:6a}). Substituting the explicit form of tensor $T^{\mu\nu}$ (and its dual one $^{\ast}T^{\mu\nu}$), Eq.\,(\ref{eq:6w}), into the equations and collecting similar terms, we lead to
\[
\begin{split}
\Bigl\{\frac{1}{2}\,\omega_{\mu\nu\,}\omega^{\hspace{0.02cm}\mu\nu} - 3\Bigl\}(S\!\hspace{0.05cm}\dot{S} - P\!\hspace{0.04cm}\dot{P}) &=
-\, \frac{1}{2}\,\omega_{\mu\nu}\!\,^{\ast}\omega^{\hspace{0.02cm}\mu\nu}(P\!\hspace{0.05cm}\dot{S} + S\!\hspace{0.04cm}\dot{P}),\\
\Bigl\{\frac{1}{2}\,\omega_{\mu\nu\,}\omega^{\hspace{0.02cm}\mu\nu} - 3\Bigl\}(P\!\hspace{0.05cm}\dot{S} + S\!\hspace{0.04cm}\dot{P})
&= +\, \frac{1}{2}\,\omega_{\mu\nu}\!\,^{\ast}\omega^{\hspace{0.02cm}\mu\nu}(S\!\hspace{0.045cm}\dot{S} - P\!\hspace{0.05cm}\dot{P}).
\end{split}
\]
This system can be presented in a more visual matrix form
\begin{align}
\begin{pmatrix}
\displaystyle\frac{1}{2}\,x - 3 & \displaystyle\frac{1}{2}\,y \\
-\displaystyle\frac{1}{2}\,y & \displaystyle\frac{1}{2}\,x - 3%
\end{pmatrix}%
\begin{pmatrix}
S\!\hspace{0.05cm}\dot{S} - P\!\hspace{0.04cm}\dot{P} \\
P\!\hspace{0.05cm}\dot{S} + S\!\hspace{0.04cm}\dot{P}%
\end{pmatrix}
&= 0,
\label{eq:6f}
\end{align}
where we have introduced the following notations
\begin{equation}
x \equiv \omega_{\mu\nu\,}\omega^{\hspace{0.02cm}\mu\nu}, \qquad y \equiv \omega_{\mu\nu}\!\,^{\ast}\omega^{\hspace{0.02cm}\mu\nu}.
\label{eq:6g}
\end{equation}
Equation (\ref{eq:6f}) has a nontrivial solution if the corresponding matrix on the left-hand side is singular. This results in the relation
\[
\Bigl(\frac{1}{2}\,x - 3\Bigr)^{2} + \frac{1}{2}\,y^2 = 0.
\]
We consider that the antisymmetric tensor $\omega_{\mu \nu}$ is real, therefore the relation above leads to the conditions
\[
x = 6, \qquad y = 0.
\]
For given representation of the tensor $\omega_{\mu \nu}$, Eq.\,(\ref{eq:6e}), the second condition holds, but the first one does not. For the case of (\ref{eq:6e}), we have
\[
x = -2.
\]
For obtaining the value $x=6$ we need to introduce a factor
\begin{equation}
\pm\, i\hspace{0.02cm} \sqrt{3}
\label{eq:6h}
\end{equation}
into the matrix $\omega_{\mu \nu}$. This violates the requirement of its reality and ipso facto of the reality of the tensor quantities $(S,V_{\mu},\!\,^{\ast}T_{\mu\nu}\hspace{0.02cm},\ldots)$ under conditions when the spinors $\psi_{\alpha}$ and $\theta_{\alpha}$ are the Majorana ones. Another way of solving this problem is to take into account only the trivial solution of system (\ref{eq:6f}), i.e. to consider that
\begin{align}
\hspace{3cm}
&S\dot{S} - P\!\hspace{0.05cm}\dot{P} = 0,&\label{eq:6j}\\
&P\!\hspace{0.04cm}\dot{S} + S\!\hspace{0.05cm}\dot{P} = 0.&\label{eq:6k}
\end{align}
The first equation can be used, for example, in the relation (\ref{eq:6d}) for its simplification. Furthermore, if we return to consideration of the right-hand side of the kinetic term (I.3.11), then the last two terms there, in view of (\ref{eq:6j}), takes the form
\[
- \frac{i}{2}\,P\!\hspace{0.04cm}\dot{P} + \frac{3\hspace{0.02cm} i}{2}\,S\!\hspace{0.05cm}\dot{S} = i\hspace{0.02cm}  P\!\hspace{0.04cm}\dot{P},
\]
i.e., the kinetic term with the pseudoscalar variable $P$ itself changes sign. Thus, for its comparison with the kinetic terms in (I.A.2) and (I.A.3) there is no need to use the constraint (I.A.13).

\section{Introducing complex tensor quantities}
\setcounter{equation}{0}
\label{section_7}

In this section we would like to consider the problem of deriving the solution of a system of algebraic equations by a slightly different way. The approach is based on introducing complex tensor variables instead of real ones. This will enable us to better understand the structure of the equations under consideration and also to clear up the question of whether the solution obtained in the previous section, is unique.\\
\indent
At first we concentrate on a system of equations of the vector type (i.e. having one external vector index), namely on the systems (\ref{eq:6q}) and (\ref{eq:6t}). Let us rewrite these equations in a more suitable form
\begin{align}
3\bigl(S\hspace{0.02cm}V^{\mu} - P\!\hspace{0.02cm}A^{\mu}\bigr) &= - V_{\nu}\hspace{0.04cm}T^{\mu\nu} - A_{\nu}\!\,^{\ast}T^{\mu\nu}, \label{eq:7q}\\
3\bigl(S\!\hspace{0.03cm}A^{\mu} + P\hspace{0.02cm}V^{\mu}\bigr) &= - A_{\nu}\hspace{0.04cm}T^{\mu\nu} - V_{\nu}\!\,^{\ast}T^{\mu\nu}, \label{eq:7w}\\
S\hspace{0.02cm}V^{\mu} + P\!\hspace{0.02cm}A^{\mu} &= - V_{\nu}\hspace{0.04cm}T^{\mu\nu} + A_{\nu}\!\,^{\ast}T^{\mu\nu}, \label{eq:7e}\\
S\!\hspace{0.035cm}A^{\mu} - P\hspace{0.02cm}V^{\mu} &= - A_{\nu}\hspace{0.04cm}T^{\mu\nu} - V_{\nu}\!\,^{\ast}T^{\mu\nu}. \label{eq:7r}
\end{align}
For the sake of convenience of future references we give here once more three independent equations of the tensor type listed in Section 3:
\begin{equation}
P\!\hspace{0.025cm}\,^{\hspace{0.02cm}\ast}T^{\mu\nu} - S\hspace{0.04cm}T^{\mu\nu} = V^{\mu}V^{\nu} + A^{\mu\!}A^{\nu},
\label{eq:7t}
\end{equation}
\begin{equation}
P\!\hspace{0.025cm}\,^{\hspace{0.02cm}\ast}T^{\mu\nu} + S\hspace{0.04cm}T^{\mu\nu} = -\,^{\ast}T^{\mu\lambda}\,^{\ast}T_{\lambda\;\;}^{\;\;\nu},
\hspace{0.45cm}
\label{eq:7y}
\end{equation}
\begin{equation}
-\hspace{0.03cm}\epsilon^{\hspace{0.025cm}\mu\nu\lambda\sigma}V_{\lambda}A_{\sigma} = V^{\mu}V^{\nu} - A^{\mu\!}A^{\nu},
\hspace{0.5cm}
\label{eq:7u}
\end{equation}
and equation of the pseudoscalar type
\begin{equation}
S\!\hspace{0.04cm}P = -\frac{1}{2}\,V_{\mu}A^{\mu}.
\label{eq:7i}
\end{equation}
\indent
Our first step is to consider equations (\ref{eq:7q}) and (\ref{eq:7w}). Multiplying the second equation by $1/i$, subtracting and adding it with the first one, we will have
\begin{equation}
\begin{split}
3\bigl(S + iP\bigr)\bigl(V^{\mu} + iA^{\mu}\bigr) &= -\hspace{0.02cm}2\hspace{0.03cm}{\rm P}^{+\hspace{0.02cm}\mu\nu\lambda\sigma}\bigl(A_{\nu}\!\,^{\ast}T_{\lambda\sigma} + V_{\nu}\hspace{0.02cm}T_{\lambda\sigma}\bigr),\\
3\bigl(S - iP\bigr)\bigl(V^{\mu} - iA^{\mu}\bigr) &= +\hspace{0.02cm}2\hspace{0.03cm}{\rm P}^{-\hspace{0.02cm}\mu\nu\lambda\sigma}\bigl(A_{\nu}\!\,^{\ast}T_{\lambda\sigma} + V_{\nu}\hspace{0.03cm}T_{\lambda\sigma}\bigr).
\end{split}
\label{eq:7o}
\end{equation}
Here, we have introduced into consideration the operators
\[
{\rm P}^{\pm\hspace{0.02cm}\mu\nu\lambda\sigma} \equiv \frac{1}{2}\,\biggl[\,\frac{1}{2}\,\bigl(\hspace{0.02cm}g^{\mu\lambda}g^{\nu\sigma} - g^{\mu\sigma\!}g^{\nu\lambda}\bigr) \pm \frac{1}{2\hspace{0.01cm}i}\,\epsilon^{\hspace{0.02cm}\mu\nu\lambda\sigma}\biggr],
\]
which project any two-form onto its self-dual $(+)$ and anti-self-dual $(-)$ two-forms \cite{Krasnov_2012}. By virtue of the specified representation of the tensor quantities
$T^{\mu\nu}$ and $\,^{\ast}T^{\mu\nu}$, Eq.\,(\ref{eq:6w}), and the definition of the projectors
${\rm P}^{\pm\hspace{0.02cm}\mu\nu\lambda\sigma}$, we have
\begin{equation}
\begin{split}
{\rm P}^{+\hspace{0.02cm}\mu\nu\lambda\sigma}\,^{\ast}T_{\lambda\sigma} &= -\hspace{0.03cm}\omega^{(+)\mu\nu\!}P\, + \,^{\ast}\omega^{(+)\mu\nu\!}S = +\hspace{0.03cm}i\bigl(S + iP\bigr)\omega^{(+)\mu\nu},\\
{\rm P}^{-\hspace{0.02cm}\mu\nu\lambda\sigma}\,^{\ast}T_{\lambda\sigma} &= -\hspace{0.03cm}\omega^{(-)\mu\nu\!}P\, + \,^{\ast}\omega^{(-)\mu\nu\!}S =
-\hspace{0.03cm}i\bigl(S - iP\bigr)\omega^{(-)\mu\nu},
\end{split}
\label{eq:7p}
\end{equation}
etc. Here,
\begin{equation}
\omega^{(\pm)\mu\nu} \equiv {\rm P}^{\pm\hspace{0.02cm}\mu\nu\lambda\sigma}\omega_{\lambda\sigma} = \frac{1}{2}\bigl[\,\omega^{\mu\nu} \mp\,i\,^{\ast}\!\omega^{\mu\nu}\bigr]
\label{eq:7a}
\end{equation}
and we have taken into account that by virtue of the definition of (anti-)self-dual two-forms the following equalities hold
\[
\,^{\ast}\omega^{(+)\mu\nu} = i\hspace{0.03cm} \omega^{(+)\mu\nu}, \qquad \,^{\ast}\omega^{(-)\mu\nu} = -i\hspace{0.03cm} \omega^{(-)\mu\nu}.
\]
From equations (\ref{eq:7o}) and projections (\ref{eq:7p}) we see that at this stage it is very natural to introduce into consideration the complex variables
\begin{equation}
\begin{split}
Z^{\mu} &\equiv V^{\mu} + \,iA^{\mu}, \quad \bar{Z}^{\mu} = V^{\mu} - iA^{\mu},\\
D &\equiv S + i\hspace{0.02cm}P, \;\qquad \bar{D} = S - i\hspace{0.02cm}P.
\end{split}
\label{eq:7s}
\end{equation}
In terms of these quantities Eqs.\,(\ref{eq:7o}) take a very compact form
\begin{equation}
\begin{split}
3D\!\hspace{0.05cm}Z_{\mu} &= 2\hspace{0.04cm} \omega^{(+)}_{\mu\nu}D\!\hspace{0.05cm}Z^{\nu}, \\
3\bar{D}\!\hspace{0.05cm}\bar{Z}_{\mu} &= 2\hspace{0.04cm} \omega^{(-)}_{\mu\nu}\bar{D}\!\hspace{0.05cm}\bar{Z}^{\nu}.
\end{split}
\label{eq:7d}
\end{equation}
Thus in such rewriting, these equations are complex conjugation of each other! It is possible to make a step forward in this direction. Note first that the equalities
\[
D^2 = \bar{D}^2 = 0
\]
are true by virtue of nilpotency. The function $D$ (or $\bar{D}$) can be canceled formally from the left- and the right-hand sides of (\ref{eq:7d}), and thus with allowance made for the previous equalities, we obtain
\begin{align}
3Z_{\mu} &= 2\hspace{0.04cm}\omega^{(+)}_{\mu\nu}Z^{\nu} + a_{\mu}D, \notag\\
3\bar{Z}_{\mu} &= 2\hspace{0.04cm}\omega^{(-)}_{\mu\nu}\bar{Z}^{\nu} + \bar{a}_{\mu}\bar{D}, \notag
\end{align}
where $a_{\mu}$ is some arbitrary (commuting) complex 4-vector.\\
\indent
Completely similar transformations for the remaining equations (\ref{eq:7e}) and (\ref{eq:7r}) lead to a system of equations
\begin{equation}
\begin{split}
\bar{D}\!\hspace{0.05cm}Z_{\mu} &= 2\hspace{0.04cm}\omega^{(-)}_{\mu\nu}\bar{D}\!\hspace{0.05cm}Z^{\nu}, \\
D\!\hspace{0.05cm}\bar{Z}_{\mu} &= 2\hspace{0.04cm}\omega^{(+)}_{\mu\nu}D\!\hspace{0.05cm}\bar{Z}^{\nu}.
\end{split}
\label{eq:7f}
\end{equation}
Here, we can also cancel the nilpotent function $\bar{D}$ (or $D$) and thereby we need to introduce into consideration another arbitrary vector $b_{\mu}$ in addition to $a_{\mu}$.\\
\indent
By analogy with (\ref{eq:6r}) we will seek a solution of a system of equations (\ref{eq:7d}) and (\ref{eq:7f}) in the form of decomposition
\begin{equation}
Z_{\mu} = \alpha_{\mu}D + \beta_{\mu}\bar{D},
\label{eq:7g}
\end{equation}
where $\alpha_{\mu}$ and $\beta_{\mu}$ are some unknown commuting (complex) vectors. Substituting (\ref{eq:7g}) into (\ref{eq:7d}) and (\ref{eq:7f}), we result in homogeneous algebraic equations for the vectors $\alpha_{\mu}$ and $\beta_{\mu}$, respectively
\begin{align}
&\bigl(\hspace{0.02cm}g_{\mu\nu} - 2\hspace{0.04cm}\omega^{(-)}_{\mu\nu}\hspace{0.02cm}\bigr)\alpha^{\nu} = 0, \label{eq:7h}\\
&\bigl(\hspace{0.02cm}3\hspace{0.02cm}g_{\mu\nu} - 2\hspace{0.04cm}\omega^{(+)}_{\mu\nu}\hspace{0.02cm}\bigr)\beta^{\nu} = 0. \label{eq:7j}
\end{align}
The condition of existence of nontrivial solutions for the equations is singularity of the corresponding matrices
\begin{align}
&\det\bigl(g_{\mu\nu} - 2\hspace{0.04cm}\omega^{(-)}_{\mu\nu}\bigr) = 0, \label{eq:7k}\\
&\det\bigl(3\hspace{0.02cm}g_{\mu\nu} - 2\hspace{0.04cm}\omega^{(+)}_{\mu\nu}\bigr) = 0. \label{eq:7l}
\end{align}
Further, for analysis of these determinants we make use of the following general formula \cite{synge_book_1956, dubrovin_book_1992}:
\begin{equation}
\det\bigl(\lambda\hspace{0.02cm}g_{\mu\nu} - {\cal F}_{\mu\nu}\bigr) = -\lambda^4 - \frac{1}{2}\hspace{0.04cm}\lambda^2 \bigl({\cal F}_{\mu\nu}{\cal F}^{\hspace{0.02cm}\mu\nu}\bigr)
+ \frac{1}{16}\bigl({\cal F}_{\mu\nu}\!\,^{\ast}\!{\cal F}^{\hspace{0.02cm}\mu\nu}\bigr)^2,
\label{eq:7z}
\end{equation}
where $\lambda$ is some eigenvalue and ${\cal F}_{\mu \nu}$ is an arbitrary antisymmetric tensor (generally speaking, it is complex). Let us consider the first condition (\ref{eq:7k}). In this special case we have
\[
{\cal F}_{\mu\nu} \equiv  2\hspace{0.04cm}\omega^{(-)}_{\mu\nu}.
\]
By virtue of anti-self-duality of the form $\omega_{\mu \nu}^{(-)}$ it follows that
\[
\,^{\ast}\!{\cal F}^{\hspace{0.02cm}\mu\nu} = -i\hspace{0.02cm}{\cal F}^{\hspace{0.02cm}\mu\nu}.
\]
In the notations (\ref{eq:6g}) the tensor contractions on the right-hand side of (\ref{eq:7z}) take a simple form
\[
\begin{split}
&{\cal F}_{\mu\nu\hspace{0.04cm}}{\cal F}^{\hspace{0.02cm}\mu\nu} = 2\hspace{0.03cm}(x + i\hspace{0.02cm}y)\equiv 2\hspace{0.02cm}z, \\
&{\cal F}_{\mu\nu}\!\,^{\ast}\!{\cal F}^{\hspace{0.02cm}\mu\nu} = -2\hspace{0.02cm}i\hspace{0.01cm}z,
\end{split}
\]
and the characteristic equation (\ref{eq:7k}), (\ref{eq:7z}) goes over into the equation
\begin{equation}
\lambda^4 + z\lambda^2 + \frac{1}{4}\,z^2 = 0.
\label{eq:7x}
\end{equation}
The discriminant of this equation relative to the variable $\lambda^2$ equals zero. Therefore, it has two, twofold degenerate, roots which are defined by the formula
\begin{equation}
\lambda^2 =  -\frac{1}{2}\,z.
\label{eq:7c}
\end{equation}
For algebraic equation (\ref{eq:7h}) we have the explicit value for $\lambda$, namely
\[
\lambda = 1,
\]
and thus the formula (\ref{eq:7c}) should be considered as a condition on the variable $z$ and finally, by virtue of the definitions (\ref{eq:6g}) as a condition on the tensor $\omega_{\mu\nu}$. In this specific case we obtain
\begin{equation}
y\,(\equiv \omega_{\mu\nu}\!\,^{\ast}\omega^{\hspace{0.02cm}\mu\nu}) =0, \qquad x\, (\equiv\omega_{\mu\nu\,}\omega^{\hspace{0.02cm}\mu\nu}) = -2.
\label{eq:7v}
\end{equation}
We have already faced with these restrictions in analysis of algebraic equations in the commutative case, Eqs.\,(\ref{eq:5u}) and (\ref{eq:5o}). The restrictions identically hold for a special choice of the antisymmetric tensors $\omega_{\mu\nu}$ and $\!\,^{\ast}\omega_{\mu\nu}$ in the presentation by means of tetrad $h_{\mu}^{(s)}$, Eq.\,(\ref{eq:6e}).\\
\indent
Let us consider the second determinant (\ref{eq:7l}). In the characteristic equation (\ref{eq:7z}) it should be set
\[
{\cal F}_{\mu\nu} \equiv  2\hspace{0.04cm}\omega^{(+)}_{\mu\nu}, \quad
\,^{\ast}\!{\cal F}^{\hspace{0.02cm}\mu\nu} = i\hspace{0.02cm}{\cal F}^{\hspace{0.02cm}\mu\nu},
\]
and in terms of quantities (\ref{eq:6g}) it takes the form
\[
\lambda^4 + \bar{z}\lambda^2 + \frac{1}{4}\,\bar{z}^2 = 0, \quad \bar{z} = x -iy.
\]
Twofold degenerate roots of this equation are now defined from
\[
\lambda^2 =  -\frac{1}{2}\,\bar{z}.
\]
By virtue of the original equation (\ref{eq:7l}), here it should be considered that
\[
\lambda = 3.
\]
This leads in turn to the following restrictions on the tensor $\omega_{\mu\nu}$:
\begin{equation}
y = 0, \qquad x = -18.
\label{eq:7b}
\end{equation}
The latter condition contradicts (\ref{eq:7v}) and thereby contradicts the choice of the representation (\ref{eq:6e}). Thus, two conditions (\ref{eq:7k}) and (\ref{eq:7l}) of the existence of nontrivial solution in the most general form (\ref{eq:7g}) contradicts each other. To overcome the difficulty we must set
\begin{equation}
\beta_{\mu} \equiv 0,
\label{eq:7n}
\end{equation}
that reduces (\ref{eq:7g}) to the form
\[
Z_{\mu} = \alpha_{\mu} D.
\]
In this case the system (\ref{eq:7d}) is identically equal to zero and we remain only with (\ref{eq:7f}). One can easy to verify that it is the same solution obtained in the previous section, Eq.\,(\ref{eq:6r}). Really, if we recall the definition of the function $Z_{\mu}$, Eq.\,(\ref{eq:7s}), and substitute (\ref{eq:6r}) instead of $V_{\mu}$ and $A_{\mu}$, then we obtain
\[
Z_{\mu} = \bigl(h^{(1)}_{\mu} +\, i\hspace{0.013cm}h^{(2)}_{\mu}\bigr) D,
\]
i.e.
\begin{equation}
\alpha_{\mu} \equiv h^{(1)}_{\mu} +\, i\hspace{0.015cm}h^{(2)}_{\mu}.
\label{eq:7m}
\end{equation}
By straightforward computation we can verify that the 4-vector $\alpha_{\mu}$ is really nontrivial zeroth mode of equation (\ref{eq:7h}).


\section{Possibility of the existence of one further solution}
\setcounter{equation}{0}
\label{section_8}

We now consider the following question:  whether it is possible to construct the second solution of algebraic system (\ref{eq:7q})\,--\,(\ref{eq:7i}), if in the decomposition (\ref{eq:7g}), instead of (\ref{eq:7n}), we set the following condition:
\[
\alpha_{\mu} \equiv 0,
\]
i.e. we seek a solution in the form
\begin{equation}
Z_{\mu} = \beta_{\mu}\bar{D}.
\label{eq:8q}
\end{equation}
In this case the second system (\ref{eq:7f}) identically vanishes and the 4-vector $\beta_{\mu}$ satisfies homogeneous equation (\ref{eq:7j}). As was shown above, this equation will have a non-trivial solution when the tensor $\omega_{\mu\nu}$ satisfies the two conditions (\ref{eq:7b}).  The first condition $\omega_{\mu\nu}\!\,^{\ast}\omega^{\hspace{0.02cm}\mu\nu}=0$ arises also in the case of commuting tensor quantities, Eq.\,(\ref{eq:5u}). In regard to the second condition $\omega_{\mu\nu}\hspace{0.03cm}\omega^{\hspace{0.02cm}\mu\nu}=-2$, in the commutative case it must be held by virtue of the equation of the pseudoscalar type, Eq.\,(\ref{eq:5i}). The overall coefficient in the tensor $\omega_{\mu\nu}$ in the presentation (\ref{eq:5a}) is chosen in such a way that the $\omega_{\mu\nu}$ satisfies Eq.\,(\ref{eq:5o}).\\
\indent
It is successful that in the anticommutative case the algebraic equations\footnote{\,Here Eq.\,(\ref{eq:6u}), on the right-hand side of which the vector quantities are contracted instead of the tensor ones, is analog of Eq.\,(\ref{eq:5i}).} do not impose such a severe restriction on the contraction $\omega_{\mu\nu}\hspace{0.03cm}\omega^{\hspace{0.02cm}\mu\nu}$. One can minimally redefine the form of antisymmetric tensor $\omega_{\mu\nu}$ with the aim of constructing a nontrivial solution of Eq.\,(\ref{eq:7j}). At first we enter the factor 3 into the initial definition (\ref{eq:6e})
\begin{equation}
\begin{split}
\omega_{\mu\nu} &= - 3\hspace{0.03cm}\epsilon_{\mu\nu\lambda\sigma} h^{(1)\lambda}h^{(2)\sigma}\!,\\
\,^{\ast}\omega_{\mu\nu} &= 3\bigl(h_{\mu}^{(1)}h_{\nu}^{(2)} - h_{\nu}^{(1)}h_{\mu}^{(2)}\bigr).
\end{split}
\label{eq:8w}
\end{equation}
Further, by analogy with (\ref{eq:7m}) we will seek a solution of Eq.\,(\ref{eq:7j}) in the following form
\begin{equation}
\beta_{\mu} = C\bigl(h^{(1)}_{\mu} +\, i\hspace{0.01cm}h^{(2)}_{\mu}\bigr),
\label{eq:8e}
\end{equation}
where $C$ is an arbitrary constant. Let us substitute (\ref{eq:8e}) into equation (\ref{eq:7j}). Taking into account the new definition (\ref{eq:8w}), we have
\[
\bigl(3\hspace{0.005cm}g^{\mu\nu} - 2\hspace{0.03cm}\omega^{(+)\mu\nu}\bigr)\beta_{\nu} = \bigl(3\hspace{0.01cm}g^{\mu\nu} - \bigl[\,\omega^{\hspace{0.02cm}\mu\nu} -\,i\,^{\ast}\!\omega^{\hspace{0.02cm}\mu\nu}\bigr]\bigr)
C\bigl(h^{(1)}_{\nu} +\, i\hspace{0.01cm}h^{(2)}_{\nu}\bigr) =
\]
\[
= 3\hspace{0.02cm}C\bigl(h^{(1)\mu} +\, i\hspace{0.01cm}h^{(2)\mu}\bigr) -
3\hspace{0.02cm}C\bigl((h^{(2)})^2h^{(1)\mu} +\, i\hspace{0.015cm}(h^{(1)})^2h^{(2)\mu}\bigr).
\]
We see that the right-hand side reduces to zero when the 4-vectors $h_{\mu}^{(1)}$ and $h_{\mu}^{(2)}$ satisfy the different normalization conditions
\begin{equation}
(h^{(1)})^2 = (h^{(2)})^2 = 1,
\label{eq:8r}
\end{equation}
i.e. these vectors must be {\it time-like} in contrast to the original ones (\ref{eq:5p}). By using the solution (\ref{eq:8q}) with (\ref{eq:8e}) it is not difficult to recover an explicit form of the real vector quantities $V_{\mu}$~and~$A_{\mu}$:
\begin{equation}
\begin{split}
V_{\mu} &= C\bigl(S\hspace{0.005cm} h^{(1)}_{\mu} + P h^{(2)}_{\mu}\bigr),\\
A_{\mu} &= C\bigl(-P h^{(1)}_{\mu} + S\hspace{0.005cm} h^{(2)}_{\mu}\bigr).
\end{split}
\label{eq:8t}
\end{equation}
By direct substitution of these expressions into a system of the ``vector'' equations (\ref{eq:7q})\,--\,(\ref{eq:7r}) with the use of (\ref{eq:6w}) for the anticommuting tensor $T_{\mu\nu}$ (where as $\omega^{\mu\nu}$ one means (\ref{eq:8w}) with the normalization (\ref{eq:8r})) we verify that this system holds (for arbitrary $C$), as it should be. Further, in equations of the ``tensor'' type (\ref{eq:7t})\,--\,(\ref{eq:7u}) two last equations hold also for arbitrary $C$. The pseudoscalar equation (\ref{eq:7i}) will be true only if we set
\begin{equation}
C^{\hspace{0.03cm}2} = 1.
\label{eq:8y}
\end{equation}
It only remains to consider Eq.\,(\ref{eq:7t}). We write out it once more
\begin{equation}
P\!\hspace{0.05cm}\,^{\hspace{0.02cm}\ast}T^{\mu\nu} - S\hspace{0.04cm}T^{\mu\nu} = V^{\mu}V^{\nu} + A^{\mu\!}A^{\nu}.
\label{eq:8u}
\end{equation}
Upon substitution of the functions  (\ref{eq:6w}) with  (\ref{eq:8w}) into the left-hand side we will have
\[
P\!\hspace{0.05cm}\,^{\hspace{0.02cm}\ast}T^{\mu\nu} - S\hspace{0.04cm}T^{\mu\nu} = -\hspace{0.01cm}2\,^{\ast}\!\omega^{\hspace{0.02cm}\mu\nu}\hspace{0.02cm}\!S\!\hspace{0.04cm}P,
\]
while on the right-hand side we have
\[
 V^{\mu}V^{\nu} + A^{\mu\!}A^{\nu} = \frac{2}{3}\,C^{\hspace{0.02cm}2\!}\;^{\ast}\!\omega^{\hspace{0.02cm}\mu\nu}\hspace{0.02cm}\!S\!
 \hspace{0.04cm}P.
 \]
The condition for the fulfillment of (\ref{eq:8u}) demands that
\begin{equation}
C^{\hspace{0.03cm}2} = -\hspace{0.01cm}3,\quad {\rm or}\quad C = \pm\hspace{0.05cm} i\sqrt{3}.
\label{eq:8i}
\end{equation}
This contradicts (\ref{eq:8y}) and, in addition, the vector variables (\ref{eq:8t}) become purely imaginary. It is interesting to note that the same factor (\ref{eq:8i}) has already arisen in Section \ref{section_6}, Eq.\,(\ref{eq:6h}). Thus, the algebraic equation of the ``tensor'' type (\ref{eq:7t}) is the only one of the total system, which does not give the possibility for the construction of a further independent solution. The reason of such preferability of the first solution (\ref{eq:6w})\,--\,(\ref{eq:6r}) is unclear for us.\\
\indent
We note also that another problem here, is the determination of an explicit form of the 4-vectors $h_{\mu}^{(1)}$ and $h_{\mu}^{(2)}$. It is easy to construct vectors satisfying (\ref{eq:8r}) from the analogous space-like vectors (\ref{ap:E1}) and (\ref{ap:E2}). For this it is sufficient to replace $S_1\rightleftharpoons C_1$ in (\ref{ap:E1}), and  $S_2\rightleftharpoons C_2$ in (\ref{ap:E2}), i.e. instead of (\ref{ap:E1}) and (\ref{ap:E2}) to take
\begin{align}
&h_{\mu}^{(1)} = \bigl( C_1, -S_1 {\bf a}^{(1)}\bigr), \notag \\
&h_{\mu}^{(2)} = \bigl( C_1C_2, -S_1C_2\hspace{0.03cm}{\bf a}^{(1)} - S_2\hspace{0.03cm}{\bf a}^{(2)}\bigr). \notag
\end{align}
However, such simple approach leads to the violation of the orthogonality condition
\[
h_{\mu}^{(1)}h^{(2)\mu} = C_2 \neq 0,
\]
and here, most likely a more subtle consideration is necessary.


\section{\bf The case of Dirac spinors $\psi_{\alpha}$ and $\theta_{\alpha}$}
\setcounter{equation}{0}
\label{section_10}

In Paper I we have concentrated on mapping the Lagrangian (I.1.4) in which the classical commuting spinor $\psi_{\alpha}$ was considered as Majorana one (and correspondingly, the auxiliary spinor $\theta_{\alpha}$ was also Majorana one). In this connection, it is worth noting that in the case of Majorana spinors the four-component formalism is not technically optimal. Here, it is more adequately to use the two-component Weyl formalism
\begin{equation}
\psi_{\rm M} =
\left(
\begin{array}{c}
\psi_{\alpha}\\
\bar{\psi}^{\dot{\alpha}}
\end{array}
\right),
\qquad
\theta_{\rm M} =
\left(
\begin{array}{c}
\theta_{\alpha}\\
\bar{\theta}^{\dot{\alpha}}
\end{array}
\right),
\label{eq:10q}
\end{equation}
where now $\alpha, \dot{\alpha} = 1, 2$ are the standard $S\!\hspace{0.04cm}U(2)$ Weyl indices. In the two-component notations the mapping (\ref{eq:5w}) can be written in the form
\begin{equation}
\hbar^{1/2}(\bar{\theta}_{\rm M}\theta_{\rm M})\psi_{\alpha} = \frac{1}{4}\hspace{0.03cm}
\Bigl\{-i\hspace{0.01cm}S\hspace{0.02cm}\theta_{\alpha} +
V_{\mu}(\sigma^{\mu})_{\alpha\dot{\alpha}\,}\bar{\theta}^{\dot{\alpha}} \,+
\,^{\ast}T_{\mu\nu}(\sigma^{\mu\nu})_{\alpha\,\,\,}^{\,\;\;\beta}\theta_{\beta} +
i\hspace{0.01cm}A_{\mu}(\sigma^{\mu})_{\alpha\dot{\alpha}\,}\bar{\theta}^{\dot{\alpha}}  -
P\hspace{0.02cm}\theta_{\alpha}\Bigr\}
\label{eq:10w}
\end{equation}
\[
\equiv
\frac{1}{4}\hspace{0.03cm}
\Bigl\{-i\hspace{0.01cm}\bar{D}\hspace{0.02cm}\theta_{\alpha} +
Z_{\mu}(\sigma^{\mu})_{\alpha\dot{\alpha}\,}\bar{\theta}^{\dot{\alpha}} \,+
\,^{\ast}T_{\mu\nu}(\sigma^{\mu\nu})_{\alpha\,\,\,}^{\,\;\;\beta}\theta_{\beta}
\Bigr\}.
\hspace{2.1cm}
\]
Here, in the notations of the textbook by Bailin and Love \cite{bailin_love_book}, we have
\[
\sigma^{\mu}\equiv (I,\,\boldsymbol{\sigma}), \quad \bar{\sigma}^{\mu}\equiv (I,\,-\boldsymbol{\sigma}),
\quad
\sigma^{\mu\nu}\equiv \frac{1}{4}\,(\sigma^{\mu}\bar{\sigma}^{\nu} - \sigma^{\nu}\bar{\sigma}^{\mu}),
\quad
\bar{\sigma}^{\mu\nu}\equiv \frac{1}{4}\,(\bar{\sigma}^{\mu}\sigma^{\nu} - \bar{\sigma}^{\nu}\sigma^{\mu}).
\]
From the last line of (\ref{eq:10w}) we see that the complex quantities $D$ and $Z_{\mu}$ introduced in Section~\ref{section_7}, Eq.\,(\ref{eq:7s}), arise naturally within the Weyl representation.\\
\indent
The term defining the interaction of a  spin with background gauge field in these variables takes the form
\[
\frac{ieg}{2}\hspace{0.03cm}\hbar\hspace{0.03cm}
(\bar{\theta}_{\rm M}\theta_{\rm M})\hspace{0.02cm}Q^{a\!}\hspace{0.01cm}
F^{\hspace{0.02cm}a}_{\mu\nu}(x)\!
\left[\left(\hspace{0.02cm}\psi^{\alpha\!}\hspace{0.01cm}(\sigma^{\mu\nu})_{\alpha\,\,\,}^{\,\;\;\beta}
\psi_{\beta}\right)
+
\bigl(\hspace{0.01cm}\bar{\psi}^{\dot{\alpha}\!}\hspace{0.01cm}
(\bar{\sigma}^{\mu\nu})_{\dot{\alpha}\,\,\,}^{\;\,\;\dot{\beta}}\bar{\psi}_{\dot{\beta}}\bigr)\right],
\]
and the kinetic term is
\[
\frac{\hbar}{i}\,(\bar{\theta}_{\rm M}\theta_{\rm M})\!\left( \psi^{\alpha}\,\frac{d\psi_{\alpha}}{d\tau} + \bar{\psi}_{\dot{\alpha}}\,\frac{d\bar{\psi}^{\dot{\alpha}}}{d\tau}\right).
\]
Thereby if an external fermion field is absent in the system, then the part of Lagrangian (I.1.4) responsible for the description of the spin degrees of freedom can be written exclusively in terms of the two-component spinor $\psi_{\alpha}$:
\[
L_{\rm spin} =
\frac{\hbar}{i}\,(\bar{\theta}_{\rm M}\theta_{\rm M})\!\left( \psi^{\alpha}\,\frac{d\psi_{\alpha}}{d\tau} -
\frac{eg}{2}\,Q^aF^{a}_{\mu\nu}(x)\!
\left(\hspace{0.02cm}\psi^{\alpha\!}\hspace{0.01cm}(\sigma^{\mu\nu})_{\alpha\,\,\,}^{\,\;\,\beta}
\psi_{\beta}\right)\!\right)
+ (\mbox{compl. conj.}),
\]
and the (one-to-one) mapping into the real tensor variables can be defined by (\ref{eq:10w}). In this case we have a set of real tensor quantities corresponding to the $n = 1$ local supersymmetric spinning particle.\\
\indent
The situation qualitatively changes in the presence of an external fermion field $\Psi_{\alpha}^i(x)$ that in the general case should be considered as a {\it Dirac} spinor. The simplest expression for the interaction of a color spinning particle with the external non-Abelian fermion field has been given in Introduction, Eq.\,(\ref{eq:5q}). It is clear that such a field inevitable violates the representation (\ref{eq:10q}) for Majorana spinors and ipso facto it is necessary from the outset to deal with the Dirac spinors $\psi_{{\rm D}\hspace{0.01cm}\alpha}$ and $\theta_{{\rm D}\hspace{0.01cm}\alpha}$. Attempt at constructing a mapping of the Dirac spinor $\psi_{{\rm D}\hspace{0.01cm}\alpha}$ results in the {\it complex} tensor quantities $(S, V_{\mu}, \,^{\ast}T_{\mu \nu}, A_{\mu}, P)$, i.e. the conditions (I.2.4) greatly facilitating our task, will not take place any more. In the general case both the tensor quantities and their complex conjugation enter into the mapping of bilinear combinations of the type $\bar{\psi}_{\rm D}\hspace{0.02cm} \hat{O}\hspace{0.02cm}\psi_{\rm D}$, where $\hat{O}$ is a differential operator or matrix, (see Eqs.\,(I.2.6), (I.2.8), (I.2.18), (I.3.2)).  In spite of the fact that a system of identities (I.C.1)\,--\,(I.C.15) has been obtained without any restrictions on the spinors $\psi_{\alpha}$ and $\theta_{\alpha}$, it is unsuitable for answer the matter which of the terms on the right-hand side of the mapping (I.2.8) and (I.3.2) are independent.\\
\indent
This raises the question as to whether it is possible to construct a system of identities similar to (I.C.1)\,--\,(I.C.15), on the left- and on right-hand sides of which the tensor quantities with their complex conjugation enter. Here, we would like to outline on a qualitative level one possible way of attacking this problem.\\
\indent
As is known, a general Dirac spinor $\psi_D$ can be always written in terms of two Majorana ones
\begin{equation}
\psi_{\rm D} = \psi_{\rm M}^{(1)} + i\hspace{0.02cm}\psi_{\rm M}^{(2)},
\label{eq:10e}
\end{equation}
where
\[
\psi_{\rm M}^{(1)} = \frac{1}{2}\,(\psi_{\rm D} + \psi_{\rm D}^{\hspace{0.02cm}c}),\qquad
\psi_{\rm M}^{(2)} = \frac{1}{2\hspace{0.02cm}i}\hspace{0.04cm}(\psi_{\rm D} - \psi_{\rm D}^{\hspace{0.02cm}c})
\]
and $\psi_{\rm D}^c$ is the charge-conjugate spinor. Such a decomposition can be performed both for the background fermion field $\Psi_{\alpha}^i(x)$ and for the spinors $\psi_{\alpha}$ and $\theta_{\alpha}$. The starting bilinear expressions, for example, the spin tensor
\begin{equation}
\frac{1}{2}\hspace{0.02cm}\hbar\hspace{0.04cm}(\bar{\theta}_{\rm D}\theta_{\rm D})
(\bar{\psi}_{\rm D}\sigma_{\mu\nu}\psi_{\rm D})
\label{eq:10r}
\end{equation}
can be presented as a product of expressions of the type
\begin{equation}
(\bar{\theta}_{\rm D}\theta_{\rm D}) = \left[\bigl(\bar{\theta}_{\rm M}^{(1)}\theta_{\rm M}^{(1)}\bigr) +
\bigl(\bar{\theta}_{\rm M}^{(2)}\theta_{\rm M}^{(2)}\bigr) \right] +
i\!\hspace{0.01cm}\left[\bigl(\bar{\theta}_{\rm M}^{(1)}\theta_{\rm M}^{(2)}\bigr) -
\bigl(\bar{\theta}_{\rm M}^{(2)}\theta_{\rm M}^{(1)}\bigr) \right]
\equiv
\bigl(\bar{\theta}_{\rm M}^{(1)}\theta_{\rm M}^{(1)}\bigr) +
\bigl(\bar{\theta}_{\rm M}^{(2)}\theta_{\rm M}^{(2)}\bigr)
\label{eq:10t}
\end{equation}
and
\[
(\bar{\psi}_{\rm D}\sigma_{\mu\nu}\psi_{\rm D}) =
\left[\bigl(\bar{\psi}_{\rm M}^{(1)}\sigma_{\mu\nu}\psi_{\rm M}^{(1)}\bigr) +
\bigl(\bar{\psi}_{\rm M}^{(2)}\sigma_{\mu\nu}\psi_{\rm M}^{(2)}\bigr) \right] +
i\!\hspace{0.01cm}\left[\bigl(\bar{\psi}_{\rm M}^{(1)}\sigma_{\mu\nu}\psi_{\rm M}^{(2)}\bigr) -
\bigl(\bar{\psi}_{\rm M}^{(2)}\sigma_{\mu\nu}\psi_{\rm M}^{(1)}\bigr) \right]
\]
\begin{equation}
\equiv
\bigl(\bar{\psi}_{\rm M}^{(1)}\sigma_{\mu\nu}\psi_{\rm M}^{(1)}\bigr) +
\bigl(\bar{\psi}_{\rm M}^{(2)}\sigma_{\mu\nu}\psi_{\rm M}^{(2)}\bigr).
\label{eq:10y}
\end{equation}
Here, on the most right-hand sides we have considered the properties of commuting and anticommuting Majorana spinors. By analogy with (I.2.1) let us consider the expansion of the spinor structure of the mixed type
\begin{equation}
\hbar^{1/2}\hspace{0.01cm}\bar{\theta}_{{\rm M}\hspace{0.01cm}\beta}^{\hspace{0.02cm}(j)}\psi_{{\rm M}\hspace{0.01cm}\alpha}^{(i)} = \frac{1}{4}\,
\Bigl\{-i\hspace{0.02cm}S^{\hspace{0.015cm}ij}\hspace{0.01cm}\delta_{\alpha\beta} + V_{\mu}^{\hspace{0.015cm}ij}(\gamma^{\mu})_{\alpha\beta} - \frac{i}{2}\,^{\ast}T_{\mu\nu}^{\hspace{0.025cm}ij}(\sigma^{\mu\nu}\gamma_{5})_{\alpha\beta} +
i\hspace{0.01cm}A_{\mu}^{ij}(\gamma^{\mu}\gamma_{5})_{\alpha\beta} + P^{\hspace{0.02cm}ij}(\gamma_{5})_{\alpha\beta}\Bigr\}
\label{eq:10u}
\end{equation}
and the expansion for the conjugate expression
\begin{equation}
\hbar^{1/2}\hspace{0.01cm}\bar{\psi}_{{\rm M}\hspace{0.01cm}\beta}^{(i)}
\theta_{{\rm M}\hspace{0.01cm}\alpha}^{(j)} = \frac{1}{4}\,
\Bigl\{i\hspace{0.02cm}S^{\hspace{0.015cm}ij}\hspace{0.01cm}\delta_{\alpha\beta} + V_{\mu}^{\hspace{0.015cm}ij}(\gamma^{\mu})_{\alpha\beta} -
\frac{i}{2}\;^{\ast}T_{\mu\nu}^{\hspace{0.025cm}ij}(\sigma^{\mu\nu}\gamma_{5})_{\alpha\beta} -
i\hspace{0.01cm}A_{\mu}^{ij}(\gamma^{\mu}\gamma_{5})_{\alpha\beta} - P^{\hspace{0.02cm}ij}(\gamma_{5})_{\alpha\beta}\Bigr\},
\label{eq:10i}
\end{equation}
where the  {\it real} anticommuting tensor variables on the right-hand side are defined as follows:
\[
\begin{split}
S^{\hspace{0.02cm}ij} \equiv i\hspace{0.02cm}\hbar^{1/2}&(\bar{\theta}^{\hspace{0.02cm}(j)}_{\rm M}\hspace{0.02cm}
\psi^{(i)}_{\rm M}),\quad\;
V_{\mu}^{\hspace{0.015cm}ij} \equiv \hbar^{1/2}(\bar{\theta}^{\hspace{0.02cm}(j)}_{\rm M}\gamma_{\mu}\hspace{0.02cm}
\psi^{(i)}_{\rm M}),\quad\;
\,^{\ast}T_{\mu\nu}^{\hspace{0.025cm}ij} \equiv i\hspace{0.02cm}\hbar^{1/2}(\bar{\theta}^{\hspace{0.02cm}(j)}_{\rm M\!}\sigma_{\mu\nu}\gamma_{5\hspace{0.02cm}}\psi^{(i)}_{\rm M}),\quad\;\\
&A_{\mu}^{ij} \equiv i\hspace{0.02cm}\hbar^{1/2}
(\bar{\theta}^{\hspace{0.02cm}(j)}_{\rm M}\gamma_{\mu}\gamma_{5\hspace{0.02cm}}\psi^{(i)}_{\rm M}),\quad\;
P^{\hspace{0.02cm}ij} \equiv \hbar^{1/2}
(\bar{\theta}^{\hspace{0.02cm}(j)}_{\rm M}\gamma_{5\hspace{0.02cm}}\psi^{(i)}_{\rm M}),\\
\end{split}
\]
$i,j = 1,2$. Multiplication of these two expansions (\ref{eq:10u}) and (\ref{eq:10i}) and the contraction of the obtained expression with $\delta_{\beta \delta} (\sigma_{\mu \nu})_{\gamma \alpha}$, yield the following generalization of the expression (I.2.10):
\begin{equation}
\hbar\hspace{0.02cm}(\bar{\theta}_{\rm M}^{\hspace{0.02cm}(j)}\theta_{\rm M}^{(l)})
(\bar{\psi}_{\rm M}^{(k)}\sigma_{\mu\nu}\psi_{\rm M}^{(i)}) = \frac{i}{4}\Bigl\{-\bigl[\hspace{0.02cm}S^{\hspace{0.02cm}ij}T_{\mu\nu}^{\hspace{0.02cm}kl} - T_{\mu\nu}^{\hspace{0.02cm}ij}\hspace{0.02cm} S^{\hspace{0.02cm}kl}\bigr]
+ \bigl[\,^{\ast}T_{\mu\nu}^{\hspace{0.025cm}ij} P^{\hspace{0.02cm}kl} -
P^{\hspace{0.025cm}ij} \,^{\ast}T_{\mu\nu}^{\hspace{0.02cm}kl}\bigr] \,+
\label{eq:10o}
\end{equation}
\[
\bigl[\hspace{0.03cm}V_{\mu}^{\hspace{0.02cm}ij}V_{\nu}^{kl} - V_{\nu}^{\hspace{0.02cm}ij}V_{\mu}^{kl}\bigr] - \bigl[\hspace{0.03cm}A_{\mu}^{ij}A_{\nu}^{kl} - A_{\nu}^{ij}A_{\mu}^{kl}\bigr] -
\epsilon_{\hspace{0.02cm}\mu\nu\;\;\;}^{\;\;\;\;\;\lambda\sigma}
\bigl[\hspace{0.03cm}V_{\lambda}^{\hspace{0.02cm}ij}A_{\sigma}^{kl} + A_{\lambda}^{ij}V_{\sigma}^{kl}\bigr] +
g^{\lambda\sigma\!}\bigl[\,^{\ast}T_{\mu\lambda}^{\hspace{0.025cm}ij}
\!\,^{\ast}T_{\sigma\nu}^{\hspace{0.02cm}kl} - \,^{\ast}T_{\nu\sigma}^{\hspace{0.025cm}ij}\!\,^{\ast}T_{\lambda\mu\,}^{\hspace{0.02cm}kl}\bigr]
\!\Bigr\}.
\]
This formula enables us to represent the expression for the spin tensor (\ref{eq:10r}) in a rather compact and obvious form which coincides essentially with the expression (I.2.10) with a slight modification. Actually, multiplying out (\ref{eq:10t}) and (\ref{eq:10y}), and taking into account (\ref{eq:10o}), we obtain
\begin{equation}
\hbar\hspace{0.02cm}(\bar{\theta}_{\rm D}\theta_{\rm D})(\bar{\psi}_{\rm D}\sigma_{\mu\nu}\psi_{\rm D}) =
\label{eq:10p}
\end{equation}
\[
=
\frac{i}{2}\,{\rm tr}\hspace{0.02cm}
\Bigl\{-\!\hspace{0.025cm}\bigl[\hspace{0.02cm}{\cal S}\hspace{0.04cm}{\cal T}_{\mu\nu}^{\hspace{0.03cm}t} + {\cal P}\,^{\ast}{\cal T}_{\mu\nu}^{\hspace{0.03cm}t}\bigr]
+\bigl[\hspace{0.02cm}{\cal V}_{\mu}{\cal V}_{\nu}^{\hspace{0.03cm}t} - {\cal A}_{\mu\hspace{0.01cm}}
{\cal A}_{\nu}^{t}\bigr] -\, \epsilon_{\mu\nu\lambda\sigma}{\cal V}^{\lambda}({\cal A}^{\sigma})^{t} +
g^{\lambda\sigma}\hspace{0.02cm}
\!\,^{\ast}{\cal T}_{\mu\lambda}\!\,^{\ast}{\cal T}_{\sigma\nu}^{\hspace{0.03cm}t}\Bigr\}.
\]
Here, we have introduced into consideration the matrices
\begin{align}
{\mathcal S} \equiv
\begin{pmatrix}
S^{11} & S^{12}\\
S^{21}& S^{22}
\end{pmatrix}
,\quad
{\mathcal V}_{\mu} \equiv
\begin{pmatrix}
V_{\mu}^{11} & V_{\mu}^{12}\\
V_{\mu}^{\hspace{0.02cm}21}& V_{\mu}^{\hspace{0.02cm}22}
\end{pmatrix}
,\quad
\,^{\ast}{\mathcal T}_{\mu\nu} \equiv
\begin{pmatrix}
\,^{\ast}T_{\mu\nu}^{11} & \,^{\ast}T_{\mu\nu}^{12}\\
\,^{\ast}T_{\mu\nu}^{\hspace{0.02cm}21}& \,^{\ast}T_{\mu\nu}^{\hspace{0.02cm}22}
\end{pmatrix}
\notag
,
\end{align}
and so on. The symbol $t$ denotes the transpose of a matrix and ${\rm tr}$ does the trace of the $2 \times 2$ matrices (the symbol ${\rm Sp}$ is used for the trace over the spinor indices). In deriving (\ref{eq:10p}) we have used the property of the transpose of a product of two matrices
\begin{equation}
(AB)^{t} = \pm\hspace{0.03cm} B^{\hspace{0.01cm}t\!\!}\hspace{0.02cm}A^{t},
\label{eq:10a}
\end{equation}
where the signs $\pm$ relate to the matrices composed of commuting or anticommuting elements, correspondingly.\\
\indent
One can obtain a system of bilinear identities which will be a generalization of the identities (I.C.1)\,--\,(I.C.15). For this purpose let us consider the product of two expansion (\ref{eq:10u})
\[
\begin{split}
\hbar\hspace{0.04cm}(\bar{\theta}_{{\rm M}\hspace{0.01cm}\beta}^{\hspace{0.02cm}(j)}
\psi_{{\rm M}\hspace{0.01cm}\alpha}^{(i)})
(\bar{\theta}_{{\rm M}\hspace{0.01cm}\delta}^{\hspace{0.02cm}(l)}\psi_{{\rm M}\hspace{0.01cm}\gamma}^{(k)})=
\frac{1}{16}\,
&\Bigl\{-i\hspace{0.02cm}S^{\hspace{0.025cm}ij}\hspace{0.01cm}\delta_{\alpha\beta} + V_{\mu}^{\hspace{0.015cm}ij}(\gamma^{\mu})_{\alpha\beta} - \frac{i}{2}\,^{\ast}T_{\mu\nu}^{\hspace{0.025cm}ij}(\sigma^{\mu\nu}\gamma_{5})_{\alpha\beta} +\,
\ldots\,\Bigr\}\\
\times\hspace{0.01cm}
&\Bigl\{-i\hspace{0.02cm}S^{\hspace{0.02cm}kl}\hspace{0.01cm}\delta_{\gamma\delta\,} + V_{\nu}^{kl}(\gamma^{\nu})_{\gamma\delta\,}  - \frac{i}{2}\,^{\ast}T_{\lambda\sigma}^{\hspace{0.02cm}kl}(\sigma^{\lambda\sigma}
\gamma_{5})_{\gamma\delta}  \,+\,
\ldots\,\Bigr\}.
\end{split}
\]
Contracting this expression with $\delta_{\beta \gamma} \delta_{\delta \alpha}$ and $\delta_{\beta \gamma} (\gamma_{\mu})_{\delta \alpha}$, we obtain correspondingly
\begin{align}
S^{\hspace{0.02cm}kj}S^{\hspace{0.025cm}il} & = \frac{1}{4}\,S^{\hspace{0.025cm}ij}S^{\hspace{0.02cm}kl} - \frac{1}{4}\,P^{\hspace{0.025cm}ij}P^{\hspace{0.02cm}kl} - \frac{1}{4}\,g^{\mu\nu}\hspace{0.01cm}V_{\mu}^{\hspace{0.02cm}ij}V_{\nu}^{kl} -
\frac{1}{4}\,g^{\mu\nu\!}A_{\mu}^{ij}A_{\nu}^{kl} + \frac{1}{8}\,
g^{\mu\lambda}g^{\nu\sigma}
\,^{\ast}T_{\mu\nu}^{\hspace{0.025cm}ij}\!\,^{\ast}T_{\lambda\sigma}^{\hspace{0.02cm}kl}, &\label{eq:10s}\\
S^{\hspace{0.02cm}kj}\hspace{0.01cm}V_{\mu}^{\hspace{0.015cm}il} &=
\frac{1}{4}\,\bigl(S^{\hspace{0.025cm}ij}V_{\mu}^{kl} +
V_{\mu}^{\hspace{0.015cm}ij}S^{\hspace{0.02cm}kl}\bigr) +
\frac{1}{4}\,\bigl(P^{\hspace{0.025cm}ij}A_{\mu}^{kl} - A_{\mu}^{ij}P^{\hspace{0.02cm}kl}\bigr) +
\frac{1}{8}\,g_{\mu\rho\,}\epsilon^{\hspace{0.01cm}\rho\nu\lambda\sigma}
\bigl(V_{\nu}^{\hspace{0.015cm}ij}\hspace{0.03cm}\!\,^{\ast}T_{\lambda\sigma}^{\hspace{0.02cm}kl} -
\,^{\ast}T_{\lambda\sigma}^{\hspace{0.025cm}ij}V_{\nu}^{kl}\bigr) \notag
\end{align}
\[
\hspace{10.2cm}
-\,\frac{1}{4}\,g^{\nu\lambda}\bigr(A_{\lambda}^{ij}\,^{\ast}T_{\mu\nu}^{\hspace{0.02cm}kl} -
\,^{\ast}T_{\mu\nu}^{\hspace{0.025cm}ij}A_{\lambda}^{kl}\bigr).
\]
These equations are a straightforward generalization of equations (I.C.1) and (I.C.2). In a similar way one can derive a generalization of the remaining equations (I.C.3)\,--\,(I.C.15). A system of bilinear identities thus obtained is rather cumbersome with large amount of variables. It is successful that by virtue of the structure of the right-hand side of (\ref{eq:10p}), we need not a complete system, but its rather special case. Contracting (\ref{eq:10s}) with $\delta^{ki} \delta^{jl}$, we obtain the following identities in the terms of the matrices introduced just above
\begin{align}
{\rm tr}\bigl({\mathcal S}{\mathcal S}^{\hspace{0.03cm}t}\bigr) =
\, &\frac{1}{4}\,{\rm tr}\bigl({\mathcal S}{\mathcal S}^{\hspace{0.03cm}t}\bigr) -
\frac{1}{4}\,{\rm tr}\bigl({\mathcal P}{\mathcal P}^{\hspace{0.03cm}t}\bigr) -
\frac{1}{4}\,g^{\mu\nu}{\rm tr}\bigl({\mathcal V}_{\mu}{\mathcal V}_{\nu}^{\hspace{0.03cm}t}\bigr) -
\frac{1}{4}\,g^{\mu\nu}{\rm tr}\bigl({\mathcal A}_{\mu}{\mathcal A}_{\nu}^{t}\bigr) +
\frac{1}{8}\,g^{\mu\lambda}g^{\nu\sigma}{\rm tr}
\bigl(\!\,^{\ast}{\mathcal T}_{\mu\nu}\!\,^{\ast}{\mathcal T}_{\lambda\sigma}^{\hspace{0.03cm}t}\bigr),\notag \\
{\rm tr}\bigl({\mathcal S}\hspace{0.02cm}
{\mathcal V}_{\mu}^{\hspace{0.03cm}t}\!\hspace{0.02cm}\hspace{0.02cm}\bigr) = \,
&\frac{1}{4}\,\bigl[\hspace{0.02cm}{\rm tr}\bigl({\mathcal S}\hspace{0.02cm}
{\mathcal V}_{\!\mu}^{\hspace{0.03cm}t}\bigr) +
{\rm tr}\bigl({\mathcal V}_{\!\mu}\hspace{0.02cm}{\mathcal S}^{\hspace{0.03cm}t}\bigr)\bigr] +
\frac{1}{4}\,\bigl[\hspace{0.02cm}{\rm tr}\bigl({\mathcal P}{\mathcal A}_{\mu}^{t}\bigr) -
{\rm tr}\bigl({\mathcal A}_{\mu}{\mathcal P}^{\hspace{0.03cm}t}\bigr)\bigr] &\label{eq:10d}\\
+\,&\frac{1}{8}\,g_{\mu\rho\,}\epsilon^{\hspace{0.02cm}\rho\nu\lambda\sigma}
\bigl[\hspace{0.02cm}{\rm tr}\bigl({\mathcal V}_{\nu}\!\,^{\ast}{\mathcal T}_{\lambda\sigma}^{\hspace{0.03cm}t}\bigr) -
{\rm tr}\bigl(\!\,^{\ast}{\mathcal T}_{\lambda\sigma}{\mathcal V}_{\nu}^{\hspace{0.03cm}t}\bigr)\bigr] -
\frac{1}{4}\,g^{\nu\lambda}
\bigl[\hspace{0.02cm}{\rm tr}\bigl({\mathcal A}_{\lambda}\!\,^{\ast}{\mathcal T}_{\mu\nu}^{\hspace{0.03cm}t}\bigr) -
{\rm tr}\bigl(\!\,^{\ast}{\mathcal T}_{\mu\nu}{\mathcal A}_{\lambda}^{t}\bigr)\bigr] \notag
\end{align}
and similar for the remaining identities. The transpose property (\ref{eq:10a}) permits to obtain easily the corresponding systems of identities in the case of commuting or anticommuting tensor variables. Analysis similar to that in Section 2 of Paper I leads to three independent relations of the form (\ref{eq:2a})\,--\,(\ref{eq:2d}) with the appropriate replacements
\[
P\,^{\ast}T_{\mu\nu} \rightarrow {\rm tr}\bigl({\mathcal P}\,^{\ast}{\mathcal T}_{\mu\nu}^{\hspace{0.03cm}t}\bigr),\quad
S\hspace{0.04cm}T^{\mu\nu} \rightarrow {\rm tr}\bigl({\mathcal S}\hspace{0.01cm}
{\mathcal T}_{\mu\nu}^{\hspace{0.03cm}t}\bigr),\quad
V_{\mu}V_{\nu} \rightarrow {\rm tr}\bigl({\mathcal V}_{\mu}{\mathcal V}_{\nu}^{\hspace{0.03cm}t}\bigr)
\]
and so on. From the preceding, we can write out at once the final expression for the tensor of spin (\ref{eq:10p}) (compare with (I.2.15))
\begin{equation}
\hbar\hspace{0.03cm}(\bar{\theta}_{\rm D}\theta_{\rm D})(\bar{\psi}_{\rm D}\sigma_{\mu\nu}\psi_{\rm D}) = -2\hspace{0.02cm}i\hspace{0.02cm}\Bigl\{{\rm tr}\bigl({\mathcal A}_{\mu}{\mathcal A}_{\nu}^{t}\bigr)  +
{\rm tr}\bigl({\mathcal S}\hspace{0.01cm}
{\mathcal T}_{\mu\nu}^{\hspace{0.03cm}t}\bigr)\!\Bigr\}.
\label{eq:10f}
\end{equation}
In doing so at the cost of fourfold increase of the number of the real tensor quantities we can construct a complete consistent description of dynamics of a color particle with half-integer spin, moving in the background non-Abelian gauge field.\\
\indent
In fact, however,  if we restrict our attention to the motion of the spinning particle only in an background gauge field, then considering the Dirac spinor $\psi_{\rm D}$ in the original Lagrangian (I.1.4) does not result in any qualitatively new consequences in the dynamical system describing the generalized Lagrangian of the type (I.A.1) with the increased number of the pseudoclassical  tensor variables $(S^{\hspace{0.02cm}ij}, V_{\mu}^{\hspace{0.02cm}ij}, ^{\ast}T_{\mu \nu}^{\hspace{0.02cm}ij}, \ldots)$. This is connected with the fact that in the Lorentz equation (I.A.10), in the equation for the color charge (I.A.9) and in the expression for the color current (I.A.12) solely the combination $S_{\mu \nu}$ enters as it was defined in Eq.\,(I.1.8). Thus, with the use of the explicit form of the right-hand side of (\ref{eq:10f}), if we determine the tensor of spin as follows:
\begin{equation}
S_{\mu\nu} \equiv -i \hspace{0.035cm}{\rm tr}\bigl({\mathcal A}_{\mu}\hspace{0.02cm}
{\mathcal A}_{\nu}^{t}\bigr),
\label{eq:10g}
\end{equation}
then all the above-mentioned dynamical equations remains unchanged and the function (\ref{eq:10g}) will obey equation (I.1.9). For the particle in the background gauge field the increase of the number of the tensor variables for the description of its spin degrees of freedom is purely effective, not leading to any dynamical consequences in its spin dynamics (at least on a classical level). Only if the system is subjected to the background {\it Dirac} fermionic field $\Psi_{\alpha}^i(x)$ we need to work with a complete set of the tensor variables $(S^{\hspace{0.02cm}ij}, V_{\mu}^{\hspace{0.02cm}ij},
\!\,^{\ast}T_{\mu \nu}^{\hspace{0.02cm}ij}, A_{\mu}^{ij}, P^{\hspace{0.02cm}ij})$.\\
\indent
Let us consider our model Lagrangian of the interaction of a color spinning particle with the non-Abelian background fermionic field $\Psi_{\rm D}^i(x)$, Eq.\,(\ref{eq:5q}). We need to define the mapping of the following expression
\begin{equation}
\hbar\hspace{0.04cm}(\bar{\theta}_{\rm D}\theta_{\rm D})
\Bigl\{\!\hspace{0.03cm}\bigl(\bar{\Psi}^{i}_{\rm D}(x)\hspace{0.03cm}\psi_{\rm D}\bigr) +
\bigl(\bar{\psi}_{\rm D}\Psi^{i}_{\rm D}(x)\bigr)\!\Bigr\}.
\label{eq:10h}
\end{equation}
The factor in front of the brace in terms of Majorana spinors is given by the expression (\ref{eq:10t}). If we will also represent the background Dirac fermion field in terms of two Majorana spinors
\[
\Psi_{\rm D}^{i} = \Psi_{\rm M}^{(1)i} + i\hspace{0.02cm}\Psi_{\rm M}^{(2)i},
\]
then, as it is not difficult to show that the following relations hold
\begin{align}
\bigl(\bar{\Psi}^{i}_{\rm D}(x)\hspace{0.03cm}\psi_{\rm D}\bigr) &=
+2\hspace{0.02cm}(\delta^{jk} + i\hspace{0.02cm}\varepsilon^{jk})
\bigl(\bar{\Psi}^{(j)i}_{\rm M}(x)\hspace{0.03cm}\psi_{\rm M}^{(k)}\bigr), \notag\\
\bigl(\bar{\psi}_{\rm D}\Psi^{i}_{\rm D}(x)\bigr) &=
-2\hspace{0.02cm}(\delta^{jk} - i\hspace{0.02cm}\varepsilon^{jk})
\bigl(\bar{\Psi}^{(j)i}_{\rm M}(x)\hspace{0.03cm}\psi_{\rm M}^{(k)}\bigr), \notag
\end{align}
where $\varepsilon^{jk}$ is the unit antisymmetric tensor (with components $\varepsilon^{12}
= - \varepsilon^{21} = 1)$ and the sum over repeated indices is implied. Taking into account these relations and (\ref{eq:10t}), the expression (\ref{eq:10h}) takes the initial form for the mapping
\[
\hbar\hspace{0.04cm}(\bar{\theta}_{\rm D}\theta_{\rm D})
\Bigl\{\!\bigl(\bar{\Psi}^{i}_{\rm D}(x)\hspace{0.03cm}\psi_{\rm D}\bigr) +
\bigl(\bar{\psi}_{\rm D}\Psi^{i}_{\rm D}(x)\bigr)\!\Bigr\} =
2\hspace{0.02cm}i\hbar\hspace{0.05cm} \varepsilon_{jk}
\bigl(\bar{\theta}_{\rm M}^{(l)}\theta_{\rm M}^{(l)}\bigr)
\bigl(\bar{\Psi}^{(j)i}_{\rm M}(x)\hspace{0.03cm}\psi_{\rm M}^{(k)}\bigr).
\]
The next step is to use the expansion of the spinor structure (\ref{eq:10u}). Contracting the expansion with the auxiliary spinor $\theta_{{\rm M}\hspace{0.01cm}\beta}^{(j)}$, we obtain the relation connecting the commuting Majorana spinor $\psi_{\rm M}^{(k)}$ with the tensor variables $(S^{\hspace{0.02cm}ij}, V_{\mu}^{\hspace{0.02cm}ij}, \!\,^{\ast}T_{\mu \nu}^{\hspace{0.025cm}ij}, A_{\mu}^{ij}, P^{\hspace{0.02cm}ij})$
\begin{equation}
\hbar^{1/2}(\bar{\theta}_{\rm M}^{(l)}\theta_{\rm M}^{(l)})\psi_{{\rm M}\hspace{0.02cm}\alpha}^{(k)} =
\label{eq:10j}
\end{equation}
\[
= \frac{1}{4}\hspace{0.02cm}
\Bigl\{-i\hspace{0.02cm}S^{\hspace{0.01cm}ks}\hspace{0.02cm}\theta_{{\rm M}\hspace{0.02cm}\alpha}^{(s)} +
V_{\mu}^{\hspace{0.01cm}ks}\bigl(\gamma^{\mu}\theta^{(s)}_{\rm M}\bigr)_{\!\alpha} -
\frac{i}{2}\,^{\ast}T_{\mu\nu}^{\hspace{0.02cm}ks}\bigl(\sigma^{\mu\nu}\gamma_{5}\theta^{(s)}_{\rm M}\bigr)_{\!\alpha} +
i\hspace{0.01cm}A_{\mu}^{ks}\bigl(\gamma^{\mu}\gamma_{5}\theta^{(s)}_{\rm M}\bigr)_{\!\alpha} +
P^{\hspace{0.015cm}ks}\bigl(\gamma_{5}\hspace{0.03cm}\theta^{(s)}_{\rm M}\bigr)_{\!\alpha\!}\Bigr\}.
\]
This expression represents a direct extension of the expansion (\ref{eq:5w}). As in the case of (\ref{eq:5w}), not all tensor variables on the right-hand side of (\ref{eq:10j}) are independent. Here, we are faced again with the problem of constructing the explicit solutions of a system of bilinear algebraic identities to which the functions
$(S^{\hspace{0.02cm}ij}, V_{\mu}^{\hspace{0.02cm}ij}, ^{\ast}T_{\mu \nu}^{\hspace{0.02cm}ij}, \ldots)$ satisfy. In contrast to the problem of the motion of a particle in an external gauge field, for the problem with an external fermion field we should analyze a complete system of algebraic equations of the form (\ref{eq:10s}) rather than the reduced system of the form (\ref{eq:10d}). The presence of additional indices $(ij)$ for the tensor variables makes the solution of the problem appreciably more difficult unlike a similar problem considered in Sections \ref{section_5} and \ref{section_6}, and it requires special consideration. The construction of a mapping of the kinetic term (I.3.1) (or more exactly (I.3.2)) for the case of Dirac spinors is also a more subtle and intricate problem. In particular, this would require an appropriate extension of a system of identities  (I.3.7),  (I.3.8) and most likely a substantial increase of the number of the tensor variables as it is seen from formulas (I.D.1) and (I.D.2) for the Majorana case.
\\
\indent
Another way to deal with the Dirac spinors $\psi_{\rm D}$ and $\theta_{\rm D}$ without recourse to the decomposition (\ref{eq:10e}) is discussed in Appendix B.


\section{\!\bf Higher-order derivative Lagrangian for spinning particle}
\setcounter{equation}{0}
\label{section_11}

In the final section we would like to consider yet another possible variant for the choice of
tetrad $h_{\mu}^{(s)}\!$. It was introduced in Section \ref{section_5} in constructing an exact solution of a system of algebraic bilinear equations. In the subsequent discussion we will essentially follow G\"ursey \cite{gursey_1955} (see also Hughes \cite{hughes_1961}). As a basic element in the definition of the tetrad we choose the 4-velocity $\dot{x}_{\mu} \equiv u_{\mu}$ of a particle (within this section we assume that the evolution parameter $\tau$ is the proper time). We set
\begin{equation}
\begin{split}
&h_{\mu}^{(0)} \equiv u_{\mu},\\
&h_{\mu}^{(1)} \equiv \rho\hspace{0.02cm}\dot{h}^{(0)}_{\mu} = \rho\hspace{0.03cm}\dot{u}_{\mu},\\
&h_{\mu}^{(2)} \equiv \sigma\bigl(\dot{h}^{(1)}_{\mu} - \rho^{-1}h^{(0)}_{\mu}\bigr)
= \sigma\bigl(\rho\hspace{0.035cm}\ddot{u}_{\mu} + \dot{\rho}\hspace{0.035cm}\dot{u}_{\mu} - \rho^{-1}u_{\mu}\bigr),\\
&h_{\mu}^{(3)}  \equiv \kappa\hspace{0.02cm}\bigl(\dot{h}^{(2)}_{\mu} + \sigma^{-1}h^{(1)}_{\mu}\bigr) ,
\end{split}
\label{eq:11q}
\end{equation}
where $\rho^{-1}, \, \sigma^{-1}$ and $\kappa^{-1}$ denote the first, second and third curvatures of the world-line, respectively. An explicit form of the first and second ones is given by the expressions
\[
\begin{split}
&\rho^{-2} = - \dot{u}_{\mu} \dot{u}^{\mu}, \\
&\sigma^{-2} = - \rho^{2\hspace{0.02cm}}\ddot{u}_{\mu} \ddot{u}^{\mu} + \rho^{-2} (1 - \dot{\rho}^2).
\end{split}
\]
A system of normals (\ref{eq:11q}) obeys the relations (\ref{eq:5p}) and thereby defines the needed tetrad. In terms of tetrad the commuting antisymmetric tensor $\!\,^{\ast}\omega_{\mu\nu}$, Eq.\,(\ref{eq:5a}),  takes the following form
\begin{equation}
\,^{\ast}\omega_{\mu\nu} = h_{\mu}^{(1)}h_{\nu}^{(2)} - h_{\nu}^{(1)}h_{\mu}^{(2)}
= \rho^{2}\sigma\hspace{0.01cm}\bigl(\dot{u}_{\mu}\ddot{u}_{\nu} -
\dot{u}_{\nu}\ddot{u}_{\mu}\bigr) -
\sigma\hspace{0.02cm}\bigl(\dot{u}_{\mu}u_{\nu} - \dot{u}_{\nu}u_{\mu}\bigr).
\label{eq:11w}
\end{equation}
Having at hand the antisymmetric tensor $\!\,^{\ast} \omega_{\mu \nu}$, we can define the following additional contributions to the original Lagrangian (I.1.4)
\begin{equation}
\hbar\hspace{0.03cm}(\bar{\theta} \theta)\,^{\ast}\omega^{\mu\nu}(\bar{\psi}\sigma_{\mu\nu}\psi),
\qquad
\hbar\hspace{0.03cm}(\bar{\theta} \theta)\,^{\ast}\omega^{\mu\nu}Q^{a\!}F^{a}_{\nu\lambda}(x)
(\bar{\psi}\sigma^{\lambda\;\;\,}_{\;\;\,\mu}\psi).
\label{eq:11e}
\end{equation}
The first expression has already been suggested in the paper\footnote{\,Instead of the $^{\ast} \omega_{\mu \nu}$, G\"ursey has considered the antisymmetric tensor $\Omega_{\mu \nu}$ in a more general form
\[
\Omega_{\mu\nu} = \rho^{-1}\bigl(h_{\mu}^{(1)}h_{\nu}^{(0)} - h_{\nu}^{(1)}h_{\mu}^{(0)}\bigr)
+
\sigma^{-1}\bigl(h_{\mu}^{(1)}h_{\nu}^{(2)} - h_{\nu}^{(1)}h_{\mu}^{(2)}\bigr)
+
\kappa^{-1}\bigl(h_{\mu}^{(2)}h_{\nu}^{(3)} - h_{\nu}^{(2)}h_{\mu}^{(3)}\bigr)
=
\]
\[
= \rho^{2}\hspace{0.01cm}\bigl(\dot{u}_{\mu}\ddot{u}_{\nu} -
\dot{u}_{\nu}\ddot{u}_{\mu}\bigr) -
\rho\hspace{0.02cm}\kappa^{-1}\epsilon_{\mu\nu\lambda\sigma}\dot{u}^{\lambda}u^{\sigma}.
\]
The tensor describes the rotation of the proper frame (tetrad) $h_{\mu}^{(s)}$ attached to a world-line.}
by G\"ursey \cite{gursey_1955} in the case of zero third curvature $\kappa^{-1}$. Under the mapping (I.2.15) the first expression (up to a numerical factor) turns to
\begin{equation}
 \rho^{2}\sigma\hspace{0.01cm}\bigl(\dot{u}_{\mu}\ddot{u}_{\nu} -
\dot{u}_{\nu}\ddot{u}_{\mu}\bigr) \xi^{\mu}\xi^{\nu} -
\sigma\hspace{0.02cm}\bigl(\dot{u}_{\mu}u_{\nu} - \dot{u}_{\nu}u_{\mu}\bigr) \xi^{\mu}\xi^{\nu} +\,\ldots\,.
\label{eq:11r}
\end{equation}
The terms of a similar type really arise in some models of the Lagrangians of a relativistic spinning particle with higher derivatives, and in particular in the model presented by Polyakov \cite{polyakov_book}. Let us consider Polyakov's approach in more detail.\\
\indent
We write out the initial functional integral in which the action is defined by the Lagrangian (I.A.1) without regard for the interacti\-on terms
\begin{equation}
\begin{split}
Z = \!\int\!{\cal D}x_{\mu}{\cal D}\xi_{\mu}{\cal D}\chi\hspace{0.02cm}{\cal D}e\hspace{0.02cm}{\cal D}\xi_{5}
\,\exp\hspace{0.02cm}\Biggl\{-\!\int\limits_{0}^{1}\!d\tau\biggl[
&-\!\displaystyle\frac{1}{2\hspace{0.02cm}e}\,\dot{x}_{\mu\hspace{0.02cm}}\dot{x}^{\mu} - \displaystyle\frac{i}{2}\,\xi_{\mu\hspace{0.035cm}}\dot{\xi}^{\mu} - \displaystyle\frac{e}{2}\,m^2\\
&+ \displaystyle\frac{i}{2\hspace{0.02cm}e}\,\chi\hspace{0.03cm}\dot{x}_{\mu\hspace{0.02cm}}\xi^{\mu} +
\displaystyle\frac{i}{2}\,\xi_{5\,}\dot{\xi_{5}} + \displaystyle\frac{i}{2}\,m\chi\hspace{0.02cm}\xi_{5}\biggr]\Biggr\}.
\end{split}
\label{eq:11t}
\end{equation}
Further we follow  Polyakov's arguments in \cite{polyakov_book} closely. Our first step is the functional integrating over $\xi_5$ according to the formula
\[
\!\int\!{\cal D}\xi_{5}\,\exp\biggl\{-\!\!\int\limits_{0}^{1}\!d\tau\biggl[\,\displaystyle\frac{i}{2}\,
\xi_{5\hspace{0.02cm}}\dot{\xi_{5}} + \displaystyle\frac{i}{2}\,m\chi\hspace{0.02cm}\xi_{5}\biggr]\biggr\}
=\,
\exp\hspace{0.02cm}\biggl\{-\frac{i\hspace{0.02cm}m^2}{\!16}\!\int\limits_{0}^{1}\!\!\int\limits_{0}^{1}
\!d\tau_{1} d\tau_{2}\;{\rm sign}(\tau_{1} - \tau_{2})\hspace{0.02cm}\chi(\tau_{1})\chi(\tau_{2})\biggr\}.
\]
The integral over the gravitino field $\chi$ is also Gaussian one. Performing the $\chi$ integration with allowance for the last equality, we obtain the following expression for
the functional integral, instead of (\ref{eq:11t}),\\
\[
\begin{split}
Z = \!\!\int\!{\cal D}x_{\mu}{\cal D}\xi_{\mu}{\cal D}e
\,\exp\hspace{0.02cm}\Biggl\{-\!\!\int\limits_{0}^{1}\!d\tau\biggl[
&-\!\displaystyle\frac{1}{2\hspace{0.02cm}e}\,\dot{x}_{\mu}\dot{x}^{\mu} - \displaystyle\frac{i}{2}\,\xi_{\mu\hspace{0.035cm}}\dot{\xi}^{\mu} - \displaystyle\frac{e}{2}\,m^2
+ \frac{i}{m^2}
\biggl(\frac{1}{e}\,(\dot{x}\cdot\xi)\biggr)\frac{d}{d\hspace{0.02cm}\tau\!}
\biggl(\frac{1}{e}\,(\dot{x}\cdot\xi)\biggr)\biggr]\\
&-\frac{\!i}{4\hspace{0.02cm}m^2}\!\int\limits_{0}^{1}\!\!\int\limits_{0}^{1}\!d\tau_{1} d\tau_{2}\;{\rm sign}(\tau_{1} - \tau_{2})\hspace{0.02cm}
\frac{d}{d\hspace{0.02cm}\tau_{1}\!}\biggl(\frac{1}{e}\,(\dot{x}\cdot\xi)\biggr)
\frac{d}{d\hspace{0.02cm}\tau_{2}\!}
\biggl(\frac{1}{e}\,(\dot{x}\cdot\xi)\biggr)\!\Biggr\}
\end{split}
\]
\begin{equation}
\begin{split}
= \!\int\!{\cal D}x_{\mu}{\cal D}\xi_{\mu}{\cal D}e
\,\exp\hspace{0.02cm}\Biggl\{-\!\!\int\limits_{0}^{1}\!d\tau\biggl[
&-\!\displaystyle\frac{1}{2\hspace{0.03cm}e}\,\dot{x}_{\mu}\dot{x}^{\mu} - \displaystyle\frac{i}{2}\,\xi_{\mu\hspace{0.035cm}}\dot{\xi}^{\mu} - \displaystyle\frac{e}{2}\,m^2\\
&-\frac{i}{2\hspace{0.03cm}m^{2}e^{2}}\;\omega_{\mu\nu}[\hspace{0.05cm}x(\tau)]\hspace{0.03cm}
\xi^{\mu}\xi^{\nu} - \frac{i}{4\hspace{0.03cm}m^{2}e^{2}}\,(\xi^{\mu}\dot{\xi}^{\nu} +
\xi^{\nu}\dot{\xi}^{\mu})\hspace{0.03cm}\dot{x}_{\mu}\dot{x}_{\nu} \biggr]\\
&\qquad\qquad\qquad\quad\, + \frac{i}{2\hspace{0.03cm}m^{2}}\,\biggl[\,\frac{1}{e}\,(\dot{x}\cdot\xi)\biggr]_{\!\tau = 1}\biggl[\,\frac{1}{e}\,(\dot{x}\cdot\xi)\biggr]_{\!\tau = 0}\,
\Biggr\}.
\end{split}
\label{eq:11y}
\end{equation}
Here, the function
\[
\omega_{\mu\nu}[\hspace{0.05cm}x(\tau)]\equiv \frac{1}{2}\,(\dot{x}_{\mu}\ddot{x}_{\nu} - \dot{x}_{\nu}\ddot{x}_{\mu}) \Bigl(= \frac{1}{2}\,(u_{\mu}\dot{u}_{\nu} - u_{\nu}\dot{u}_{\mu})\Bigr)
\]
was introduced by Polyakov. It is the rotation of the tangent vector to the trajectory. An expression similar to (\ref{eq:11y}) was also considered in the different context in works \cite{gauntlett_1990} (see also \cite{galvao_1980}).\\
\indent
According to the obtained expression (\ref{eq:11y}), when the boundary term is dropped, we can choose as the Lagran\-gian for a spinning particle the following expression:
\begin{equation}
\begin{split}
L=
&-\!\displaystyle\frac{1}{2\hspace{0.02cm}e}\,\dot{x}_{\mu\hspace{0.03cm}}\dot{x}^{\mu} - \displaystyle\frac{e}{2}\,m^2\\
&- \displaystyle\frac{i}{2}\;\xi_{\mu\hspace{0.03cm}}\dot{\xi}^{\mu} - \frac{i}{4\hspace{0.03cm}m^{2}e^{2}}\,
\Bigl\{\!\hspace{0.02cm}(\dot{x}_{\mu}\ddot{x}_{\nu} - \dot{x}_{\nu}\ddot{x}_{\mu})\hspace{0.03cm}\xi^{\mu}\xi^{\nu} +\,
(\xi^{\mu}\dot{\xi}^{\nu} +\, \xi^{\nu}\dot{\xi}^{\mu})\hspace{0.03cm}\dot{x}_{\mu}\dot{x}_{\nu}\!\Bigr\} +\, \ldots\,.
\end{split}
\label{eq:11u}
\end{equation}
Formally, it is obviously independent of the pseudoscalar $\xi_5$ and the gravitino $\chi$. Nevertheless, the Lagrangian (\ref{eq:11u}) is still SUSY-invariant\footnote{\,The residual supersym\-metric transformation is
\[
\delta{x}_{\mu}=i\hspace{0.02cm}\alpha\hspace{0.02cm}\xi_{\mu},\quad
\delta{e}=\hspace{0.02cm}i\hspace{0.02cm}\alpha\,\frac{\!\!2}{m^2}\,\frac{d}{d\tau\!}\biggl(\frac{1}{e}\,
(\dot{x}\cdot\xi)\biggr),\quad
\delta{\xi}_{\mu}= -\hspace{0.02cm}\alpha\hspace{0.03cm}\dot{x}_{\mu}\frac{1}{e}
\, +\, i\hspace{0.02cm}\alpha\hspace{0.03cm}\xi_{\mu}\,\frac{\!1}{e\hspace{0.02cm}m^2}\,\frac{d}{d\tau\!}
\biggl(\frac{1}{e}\,(\dot{x}\cdot\xi)\biggr).
\]} (up to the total derivative) and depends on the 4-acceleration as well as the 4-velocity of the particle. We see that the second term on the right-hand side of (\ref{eq:11r}) exactly coincides with the first term in braces (\ref{eq:11u}). However, the expression (\ref{eq:11r}) contains one further contribution with higher (third)  derivative with respect to $\tau$. Its physical interpretation is not clear.\\
\indent
For a spinless particle in the $D$-dimensional space-time the Lagrangian with the third order derivatives of variable $x^{\mu}(\tau)$ (the relativistic particle with curvature and torsion) has been considered in the papers by  Plyushchay \cite{plyushchay_1989, plyushchay_1990, plyushchay_1991} and Nesterenko \cite{nesterenko_1991}. Inclusion into the Lagrangian of higher derivatives with respect to the proper time of dynamical variables entails the account of additional degrees of freedom. As a result, it turns out well in this approach to describe particles with a nonzero spin without introducing additional variables (in particular, anticommuting Grassmann ones).
The approach suggested in this section based on introducing tetrad (\ref{eq:11q}) and additional contributions (\ref{eq:11e}) with the mapping (\ref{eq:11r}), can be considered as a ``hybrid'' approach to the classical description of the spin degrees of freedom simultaneously involving both the Grassmann-odd variable $\xi^{\mu}(\tau)$ and the position variable $x^{\mu}(\tau)$ with higher derivatives.\\
\indent
In closing this section we note that the idea of considering the higher-derivative Lagrangians for a point classical particle with the spin (with the proper rotation) is not new and originated in the classical papers by Frenkel  \cite{frenkel_1926} and Thomas \cite{thomas_1927}. Actually, the expressions of the type (\ref{eq:11r}) are already contained in the above-mentioned papers if instead of the variable $\xi_{\mu}$ in (\ref{eq:11r}) one introduces the tensor of spin $S_{\mu \nu}$ by the rule (I.1.8). Besides the second contribution in (\ref{eq:11e}) for the case of background Abelian gauge field has also been given in \cite{thomas_1927} (Eq.\,(9.8)).


\section{Conclusion}
\setcounter{equation}{0}
\label{section_12}

In this paper we have presented further analysis of the interaction of a classical color spinning particle with background non-Abelian fermionic field. Here, we confined close attention to the spin sector of the interaction. An explicit form of the interaction terms with an external {\it Majorana} fermion field $\Psi_{{\rm M}\hspace{0.01cm} \alpha}^i(x)$ in terms of the real Grassmann-odd current variables
$(S, V_{\mu}, \!\,^{\ast}T_{\mu\nu}, A_{\mu}, P)$ can be obtained by substituting the expression (\ref{eq:5w}) in the interaction Lagrangian (\ref{eq:5q}). In the particular case the variables $V_{\mu}, A_{\mu}$ and $\!\,^{\ast}T_{\mu \nu}$ are expressed  through two independent those $S, P$ and tetrad $h_{\mu}^{(s)}$ with the aid of the relations (\ref{eq:6r}) and (\ref{eq:6w}).  An explicit form of the tetrad can be chosen in any  representation (\ref{ap:E1})\,--\,(\ref{ap:E4}), (\ref{eq:5d}) or (\ref{eq:11q}). Further, in Section \ref{section_10} we suggested a way to extend the above result to a more general case of an external {\it Dirac} fermion field $\Psi_{\!{\rm D}\hspace{0.01cm} \alpha}^i(x)$. It was shown that to cover this case one needs to introduce into consideration a similar tensor set $({\cal S}, {\cal V}_{\mu}, \!\,^{\ast} {\cal T}_{\mu \nu}, {\cal A}_{\mu}, {\cal P})$, where each of these variables represents a $2\times 2$ matrix consisting of real components. Unfortunately, an important problem of defining relations between these matrix variables has remained unsolved.\\
\indent
There is a further point to be made here. Throughout this work and Paper I we have dealt with the mapping of the (pseudo)classical models. Naturally, a more deep and principle question is connected with the construction of a mapping between the dynamical systems obtained after quantization of these classical models. Is it possible to construct such a (one-to-one) mapping of these quantized models and to what extent it will be complete? This problem is worth a careful look. The results reported in this paper are put forward as a classical starting point for the subsequent analysis of mapping the quantized systems.

\section*{\bf Acknowledgments}

 Authors thank M. Pav{\v s}i{\v c} and M.S. Plyushchay which have drawn our attention to the papers closely related to the subject of the given work. This work was supported in part by the grant of the President of Russian Federation for the support of the leading scientific schools (NSh-3003.2014.2, NSh-5007.2014.9).



\begin{appendices}
\numberwithin{equation}{section}


\section{Parameter representations of triad and tetrad}
\label{appendix_E}
\setcounter{equation}{0}

In this appendix we give an explicit form of the parameter representations of orthogonal triad and tetrad \cite{synge_book_1956, takahashi_1983}. The orthogonal triad $({\bf a}^{(1)}, {\bf a}^{(2)}, {\bf a}^{(3)})$ satisfying the orthonormality relation
\[
{\bf a}^{(i)\!}\cdot{\bf a}^{(j)} = \delta^{ij}, \quad i,j = 1,2,\,3
\]
and the completeness relation
\[
a^{(k)\!}_{i} a^{(k)}_{j} = \delta_{ij}\hspace{0.02cm},
\]
can be expressed, for example, in terms of three Euler angles $\alpha,\,\beta$ and $\gamma$
\begin{align}
&{\bf a}^{(1)} =
\begin{pmatrix}
\cos\!\hspace{0.03cm}\alpha\hspace{0.04cm} \sin\!\hspace{0.03cm}\beta \\
\sin\!\hspace{0.03cm}\alpha\hspace{0.04cm} \sin\!\hspace{0.03cm}\beta \\
\cos\beta%
\end{pmatrix}
,
\notag\\
&{\bf a}^{(2)} =
\begin{pmatrix}
\cos\!\hspace{0.03cm}\alpha\hspace{0.04cm} \cos\!\hspace{0.03cm}\beta \hspace{0.04cm}\cos\!\hspace{0.03cm}\gamma
-
\sin\!\hspace{0.03cm}\alpha \hspace{0.04cm} \sin\!\hspace{0.03cm}\gamma \\
\sin\!\hspace{0.03cm}\alpha\hspace{0.04cm} \cos\!\hspace{0.03cm}\beta \hspace{0.04cm}\cos\!\hspace{0.03cm}\gamma
+
\cos\!\hspace{0.03cm}\alpha \hspace{0.04cm} \sin\!\hspace{0.03cm}\gamma\\
-\sin\!\hspace{0.03cm}\beta   \hspace{0.04cm} \cos\!\hspace{0.03cm}\gamma%
\end{pmatrix}
,
\notag\\
&{\bf a}^{(3)} =
\begin{pmatrix}
-\cos\!\hspace{0.03cm}\alpha\hspace{0.04cm} \cos\!\hspace{0.03cm}\beta \hspace{0.04cm}\sin\!\hspace{0.03cm}\gamma
-
\sin\!\hspace{0.03cm}\alpha \hspace{0.04cm} \cos\!\hspace{0.03cm}\gamma\\
-\sin\!\hspace{0.03cm}\alpha\hspace{0.04cm} \cos\!\hspace{0.03cm}\beta \hspace{0.04cm}\sin\!\hspace{0.03cm}\gamma
+
\cos\!\hspace{0.03cm}\alpha \hspace{0.04cm} \cos\!\hspace{0.03cm}\gamma\\
\sin\!\hspace{0.03cm}\beta   \hspace{0.04cm} \sin\!\hspace{0.03cm}\gamma%
\end{pmatrix}
.
\notag
\end{align}
As is known the components of these vectors constitute the rotation matrix $(a_{ij}) \equiv (a_i^{(j)})$ in the three-dimensional Euclidean space \cite{varshalovich_book_1988}. The other parameterizations of the triad are also possible: in terms of the angles $\theta, \phi$ which describe the direction of rotation axes  and the rotation angle $\omega$ or in terms of the Cayley-Klein parameters \cite{varshalovich_book_1988, goldstein_book_1980}.\\
\indent
Further, the orthogonal tetrad $h_{\mu}^{(s)},\,s=0, 1, 2, 3$ in Minkowski space is subject to the orthonormality relation
\[
h_{\mu}^{(s)}h^{(s^{\prime})\mu} = g^{ss^{\prime}} = {\rm diag}(1,-1,-1,-1),
\]
and the completeness relation
\[
h_{\mu}^{(s)}h_{\nu}^{(s^{\prime})}g_{ss^{\prime}} = g_{\mu\nu} = {\rm diag}(1,-1,-1,-1)
\]
can be parameterized as follows
\begin{align}
&h_{\mu}^{(1)} = \bigl( S_1, -C_1 {\bf a}^{(1)}\bigr), \label{ap:E1}\\
&h_{\mu}^{(2)} = \bigl( C_1S_2, -S_1S_2\hspace{0.03cm}{\bf a}^{(1)} - C_2\hspace{0.03cm}{\bf a}^{(2)}\bigr), \label{ap:E2}\\
&h_{\mu}^{(3)} = \bigl( C_1C_2S_3, -S_1C_2S_3\hspace{0.03cm}{\bf a}^{(1)} -
S_2S_3\hspace{0.03cm}{\bf a}^{(2)} - C_3\hspace{0.03cm}{\bf a}^{(3)}\bigr), \label{ap:E3}\\
&h_{\mu}^{(0)} = \bigl( C_1C_2C_3, -S_1C_2C_3\hspace{0.03cm}{\bf a}^{(1)} -
S_2C_3\hspace{0.03cm}{\bf a}^{(2)} - S_3\hspace{0.03cm}{\bf a}^{(3)}\bigr), \label{ap:E4}
\end{align}
where $C_i \equiv \cosh \chi_i,\,S_i \equiv \sinh \chi_i,\, i= 1, 2,\hspace{0.03cm}3$; $\chi_i$ are the Eulerian pseudoangles for the Lorentz boost. The 4-vectors $h_{\mu}^{(k)} (k= 1, 2, 3)$ are space-like, whereas $h_{\mu}^{(0)}$ is time-like. Similar to the triad ${\bf a}^{(i)}$ the components of the tetrad $h_{\mu}^{(s)}$ constitute the rotation matrix in Minkowski space \cite{synge_book_1956}. Here, the other parameterizations of the 4-vectors $h_{\mu}^{(s)}$ are also permissible. One of them is mentioned at the end of Section \ref{section_5}.


\section{An extended system of bilinear identities}
\label{appendix_F}
\setcounter{equation}{0}

Here we represent another way to deal with the Dirac spinors $\psi_{\rm D}$ and $\theta_{\rm D}$ without recourse to their decomposition in terms of Majorana spinors $\psi_{\rm M}^{(i)}$ and
$\theta_{\rm M}^{(i)}$, $i = 1, 2$, Eq.\,(\ref{eq:10e}). For this purpose we shall use our basic formula of a product of the two expansions (I.2.1) and (I.2.2), namely Eq.\,(I.2.5). In addition to the expansions (I.2.1) and (I.2.2) we need also the ones of the spinor structures\footnote{\,For simplicity of notations we shall drop the symbol ${\rm D}$ for the Dirac spinors.} $\hbar^{1/2} \bar{\psi}_{\beta} \psi_{\alpha}$ and $\hbar^{1/2} \bar{\theta}_{\beta} \theta_{\alpha}$, correspondingly:
\begin{align}
\hbar^{1/2}\bar{\psi}_{\beta}\psi_{\alpha} &= \frac{1}{4}\,
\Bigl\{S_{\psi}\hspace{0.02cm}\delta_{\alpha\beta} +
V^{\mu}_{\psi}(\gamma_{\mu})_{\alpha\beta} - \frac{i}{2}\;^{\ast}T^{\mu\nu}_{\psi}(\sigma_{\mu\nu}\gamma_{5})_{\alpha\beta} -
A^{\mu}_{\psi}(\gamma_{\mu}\gamma_{5})_{\alpha\beta} -
iP_{\psi}(\gamma_{5})_{\alpha\beta}\Bigr\}, \label{ap:F1}\\
\hbar^{1/2\,}\bar{\theta}_{\beta}\hspace{0.03cm}\theta_{\alpha} &= \frac{1}{4}\,
\Bigl\{S_{\theta}\hspace{0.02cm}\delta_{\alpha\beta} +
V^{\mu}_{\theta}(\gamma_{\mu})_{\alpha\beta} - \frac{i}{2}\;^{\ast}T^{\mu\nu}_{\theta}(\sigma_{\mu\nu}\gamma_{5})_{\alpha\beta} -
A^{\mu}_{\theta}(\gamma_{\mu}\gamma_{5})_{\alpha\beta} -
iP_{\theta}(\gamma_{5})_{\alpha\beta}\Bigr\}. \label{ap:F2}
\end{align}
Here, the {\it real commuting} tensor variables on the right-hand side are defined as follows
\begin{equation}
\begin{split}
S_{\psi} \equiv \hspace{0.02cm}\hbar^{1/2}&(\bar{\psi}\hspace{0.02cm}\psi),\quad\;
V^{\mu}_{\psi} \equiv \hbar^{1/2}(\bar{\psi}\gamma^{\mu}\hspace{0.02cm}\psi),\quad\;
\,^{\ast}T^{\mu\nu}_{\psi} \equiv i\hspace{0.02cm}\hbar^{1/2}(\bar{\psi}\sigma^{\mu\nu}\gamma_{5\hspace{0.02cm}}\psi),\quad\;\\
&A^{\mu}_{\psi} \equiv \hspace{0.02cm}\hbar^{1/2}(\bar{\psi}\hspace{0.02cm}\gamma^{\mu}\gamma_{5\hspace{0.02cm}}
\psi),\quad\;
P_{\psi} \equiv i\hspace{0.01cm}\hbar^{1/2}(\bar{\psi}\hspace{0.02cm}\gamma_{5\hspace{0.02cm}}\psi),\\
\end{split}
\label{ap:F3}
\end{equation}
and similarly for $(S_{\theta}, V_{\theta}^{\mu}, ^{\ast}T_{\theta}^{\mu \nu}, A_{\theta}^{\mu}, P_{\theta})$ with the replacement $\psi \rightarrow \theta$. Recall that the latter tensor variables
are nilpotent. In particular, in terms of these variables the conditions under which the spinors $\psi_{\alpha}$ and $\theta_{\alpha}$ are Majorana ones, Eq.\,(I.2.7), take the form
\begin{equation}
S_{\psi} = A^{\mu}_{\psi} = P_{\psi}  = 0,\quad V^{\mu}_{\theta} = \,^{\ast}T^{\mu\nu}_{\theta} = 0.
\label{ap:F4}
\end{equation}
\indent
We find the first system of identities from the expression (I.2.5) by contracting the left- and right-hand sides with every possible combinations of the type
\begin{equation}
\delta_{\beta \delta} \delta_{\gamma \alpha}, \quad
\delta_{\beta \delta} (\gamma_5)_{\gamma \alpha}, \quad
(\gamma_5)_{\beta \delta} \delta_{\gamma \alpha}, \quad
\delta_{\beta \delta} (\gamma_{\mu})_{\gamma \alpha}, \quad
 (\gamma_{\mu})_{\beta \delta} \delta_{\gamma \alpha},
\label{ap:F5}
\end{equation}
etc. Equations (I.2.6) , (I.2.8)  and (I.2.18) are just a special case of these contractions. In terms of quantities (\ref{ap:F3}) equations (I.2.6) and (I.2.18) become
\begin{flalign}
\hspace{0.5cm}
&S_{\theta}S_{\psi}= \frac{1}{4}\hspace{0.03cm}\Bigl\{SS^{\hspace{0.02cm}\ast} + V_{\mu}(V^{\mu})^{\ast} - \frac{1}{2}\,^{\ast}T_{\mu\nu}(\!\,^{\ast}T^{\mu\nu})^{\ast} -
A_{\mu}(A^{\mu})^{\ast} - P\!\hspace{0.04cm}P^{\hspace{0.02cm}\ast}\Bigr\},  &\label{ap:F6}\\
&S_{\theta}V^{\mu}_{\psi} = \frac{i}{4}\hspace{0.03cm}\Bigl\{
-\hspace{0.02cm}\!\bigl[\hspace{0.02cm}S\hspace{0.02cm}(V^{\mu})^{\ast} - V^{\mu}S^{\ast}\bigr] + \bigl[\hspace{0.02cm}P\hspace{0.02cm}(A^{\mu})^{\ast} - A^{\mu\!}\hspace{0.02cm}P^{\ast}\bigr]
- \frac{1}{2}\,\epsilon^{\hspace{0.02cm}\mu\nu\lambda\sigma}
\bigl[\hspace{0.02cm}V_{\nu}(\!\,^{\ast}T_{\lambda\sigma})^{\ast} - \,^{\ast}T_{\lambda\sigma}(V_{\nu})^{\ast}\bigr]
&\notag
\end{flalign}
\[
\hspace{10.65cm}
+ \,\bigl[\hspace{0.02cm}A_{\nu}(\!\,^{\ast}T^{\mu\nu})^{\ast} - \,^{\ast}T^{\mu\nu}(A_{\nu})^{\ast}\bigr]\!\Bigr\},
\]
etc. Further, by analogy with (I.2.5) we define a product of the expansions (\ref{ap:F1}) and (\ref{ap:F2})
\[
\hbar\hspace{0.04cm}(\bar{\psi}_{\beta}\psi_{\alpha})(\bar{\theta}_{\gamma}\theta_{\delta}) =
\]
\[
\begin{split}
 \frac{1}{16}\,
&\Bigl\{S_{\psi}\hspace{0.02cm}\delta_{\alpha\beta} +
V^{\mu}_{\psi}(\gamma_{\mu})_{\alpha\beta} - \frac{i}{2}\;^{\ast}T^{\mu\nu}_{\psi}(\sigma_{\mu\nu}\gamma_{5})_{\alpha\beta} -
A^{\mu}_{\psi}\hspace{0.01cm}(\gamma_{\mu}\gamma_{5})_{\alpha\beta} -
iP_{\psi}(\gamma_{5})_{\alpha\beta}\Bigr\}\\
\times\hspace{0.02cm}
&\Bigl\{S_{\theta}\hspace{0.02cm}\delta_{\delta\gamma} +
V^{\nu}_{\theta}(\gamma_{\nu})_{\delta\gamma} - \frac{i}{2}\;^{\ast}T^{\lambda\sigma}_{\theta}(\sigma_{\lambda\sigma}\gamma_{5})_{\delta\gamma} -
A^{\nu}_{\theta}(\gamma_{\nu}\gamma_{5})_{\delta\gamma} -
iP_{\theta}(\gamma_{5})_{\delta\gamma}\Bigr\}.
\end{split}
\]
From this expression making use of the crossed contractions with the spinor structures (\ref{ap:F5}) we define the second system of identities
\begin{flalign}
\hspace{0.5cm}
&S\!\hspace{0.05cm}S^{\ast\,} = \frac{1}{4}\hspace{0.03cm}
\Bigl\{-S_{\theta}S_{\psi} - g_{\mu\nu}V_{\theta}^{\mu}V_{\psi}^{\nu}  + \frac{1}{2}\hspace{0.01cm}\,^{\ast}T_{\theta}^{\mu\nu}\,^{\ast}T_{\psi\hspace{0.01cm}\mu\nu} +
g_{\mu\nu}A_{\theta}^{\mu}A_{\psi}^{\nu}
+ P_{\theta}P_{\psi}\! \Bigr\}, &\notag\\
&S\!\hspace{0.05cm}P^{\ast} = \frac{1}{4}\hspace{0.03cm}
\Bigl\{\bigl(P_{\theta}S_{\psi} + P_{\psi}S_{\theta}\bigr) +
i\hspace{0.02cm} g_{\mu\nu}\bigl(V_{\theta}^{\mu}A_{\psi}^{\nu} - V_{\psi}^{\mu}A_{\theta}^{\nu}\bigr) -
\frac{1}{4}\,\epsilon_{\hspace{0.02cm}\mu\nu\lambda\sigma}\!\,^{\ast}T^{\mu\nu}_{\theta}
\,^{\ast}T^{\lambda\sigma}_{\psi}\!\Bigr\}, &\label{ap:F7}\\
&S\hspace{0.015cm}V_{\mu}^{\ast} = \frac{1}{4}\hspace{0.03cm}
\Bigl\{-i\hspace{0.02cm}g_{\mu\nu}\bigl(
V_{\theta}^{\nu}\hspace{0.01cm}S_{\psi} +
V_{\psi}^{\nu}\hspace{0.01cm}S_{\theta}\bigr)
+
g_{\mu\nu}\bigl(A_{\theta}^{\nu}\hspace{0.01cm}P_{\psi} - A_{\psi}^{\nu}\hspace{0.01cm}\hspace{0.01cm}P_{\theta}\bigr)
-
\frac{1}{2}\,\epsilon_{\hspace{0.02cm}\mu\nu\lambda\sigma}
\bigl(V^{\nu}_{\theta}\hspace{0.01cm}\,^{\ast}T^{\lambda\sigma}_{\psi}
- V^{\nu}_{\psi}\hspace{0.01cm}\,^{\ast}T^{\lambda\sigma}_{\theta}\bigr)
 &\notag
\end{flalign}
\[
\hspace{10.8cm}
+\, i\hspace{0.02cm}\bigl(\!\,^{\ast}T_{\theta\hspace{0.01cm}\mu\nu}\hspace{0.01cm}A^{\nu}_{\psi}
+
\!\,^{\ast}T_{\psi\hspace{0.01cm}\mu\nu}\hspace{0.01cm}A^{\nu}_{\theta}\bigr)\!
\Bigr\},
\]
etc. The corresponding systems of identities for the tensor sets $(S_{\psi}, V_{\psi}^{\mu},
\!\,^{\ast}T_{\psi}^{\mu \nu}, A_{\psi}^{\mu}, P_{\psi})$ and $(S_{\theta}, V_{\theta}^{\mu}, \!\,^{\ast}T_{\theta}^{\mu \nu}, A_{\theta}^{\mu}, P_{\theta})$ must be added to the obtained systems of bilinear equations (\ref{ap:F6}) and (\ref{ap:F7}). These systems follow from the expansions (\ref{ap:F1}) and (\ref{ap:F2}). This enables us to close a system of bilinear identities completely. In an important special case of Majorana spinors when the equalities (I.2.4) and (\ref{ap:F4}) hold, we have to reproduce the system (I.C.1)\,--\,(I.C.15).\\
\indent
The systems of identities (\ref{ap:F6}), (\ref{ap:F7}), $\ldots$ define in an {\it implicit} form the desired relations between the various functions $S\!\hspace{0.05cm}S^{\ast\!},\, S\!\hspace{0.05cm}P^{\ast\!},\, S\hspace{0.015cm}V_{\mu}^{\ast},\, \ldots$ through auxiliary tensor sets $(S_{\psi}, V_{\psi}^{\mu}, ^{\ast}T_{\psi}^{\mu \nu}, \ldots)$ and $(S_{\theta}, V_{\theta}^{\mu}, ^{\ast}T_{\theta}^{\mu \nu}, \ldots)$. Such a way of solving the problem is also rather cumbersome. However, in reality we need only the relationships between definite combinations of these functions (by virtue of the explicit form of the right-hand side of the expression (I.2.8)), namely
\[
S\hspace{0.01cm}(\hspace{0.02cm}T^{\mu\nu})^{\ast} - T^{\mu\nu\!} S^{\hspace{0.02cm}\ast},
\quad
\,^{\ast}T^{\mu\nu\!} P^{\hspace{0.02cm}\ast} - P\hspace{0.01cm} (\!\,^{\ast}T^{\mu\nu})^{\ast},
\quad
V^{\mu}(V^{\nu})^{\ast} - V^{\nu}(V^{\mu})^{\ast},
\quad
A^{\mu\!}\hspace{0.01cm}(A^{\nu})^{\ast} - A^{\nu\!}\hspace{0.01cm}(A^{\mu})^{\ast},
\]
\[
\epsilon^{\hspace{0.02cm}\mu\nu\lambda\sigma}\bigl[\hspace{0.03cm}V_{\lambda}(A_{\sigma})^{\ast} + A_{\lambda}(V_{\sigma})^{\ast}\bigr],
\quad
\,^{\ast}T^{\mu\lambda}(\!\,^{\ast}T_{\lambda\;\;}^{\;\;\nu})^{\ast} - \,^{\ast}T^{\nu\lambda}(\!\,^{\ast}T_{\lambda\;\;}^{\;\;\mu})^{\ast}.
\]
This circumstance should essentially facilitate the problem of a search for the independent tensor combinations.\\

\end{appendices}

%
%


\end{document}